\documentclass[aps,prl,reprint,groupedaddress]{revtex4-1}
\usepackage{epsf}
\usepackage{epsfig}
\usepackage{graphicx}
\usepackage{color}
\usepackage{siunitx}
\usepackage{epstopdf}
\usepackage{amsmath}
\begin{document}
\title {\large
Holographic Angular Streaking of Electrons and the Wigner-Time Delay}
\author{S. Eckart}
\email{eckart@atom.uni-frankfurt.de}
\affiliation{Institut f\"ur Kernphysik, Goethe-Universit\"at, Max-von-Laue-Str. 1, 60438 Frankfurt, Germany}
\date{\today}
\begin{abstract}
For a circularly polarized single-color field at a central frequency of $2\omega$ the final electron momentum distribution upon strong field ionization does not carry any information about the phase of the initial momentum distribution. Adding a weak, co-rotating, circularly polarized field at a central frequency of $\omega$ gives rise to a sub-cycle interference pattern (holographic angular streaking of electrons (HASE)). This interference pattern allows for the retrieval of the derivative of the phase of the initial momentum distribution after tunneling $\phi^{\prime}_{\mathrm{off}}(p_i)$. A trajectory-based semi-classical model (HASE model) is introduced which links the experimentally accessible quantities to $\phi^{\prime}_{\mathrm{off}}(p_i)$. It is shown that a change in $\phi^{\prime}_{\mathrm{off}}$ is equivalent to a displacement in position space $\Delta x$ of the initial wave packet after tunneling. This offset in position space allows for an intuitive interpretation of the Wigner time delay $\Delta \tau_W$ in strong field ionization for circularly polarized single-color fields. The influence of Coulomb interaction after tunneling is investigated quantitatively.
\end{abstract}
\maketitle
\section{I. Introduction}
The appearance of a comb of discrete peaks in the energy distributions of electrons upon strong field ionization \cite{Keldysh1965} of single atoms or molecules is well-known as above threshold ionization (ATI) \cite{voronov1966many,Freeman1987,Misha2005}. This quantization of energy can be interpreted as a consequence of energy conservation and the finite bandwith of the incident photons \cite{Einstein1905}. Alternatively, the time-dependent electric field of the incident light can be considered: for a light pulse with multiple cycles, the periodic release of electron wave packets (grating in the time-domain) gives rise to equally spaced interference fringes in energy-space. Thus, the periodicity of the light's time-dependent electric field and the discrete value for the photon energy are two sides of the same coin \cite{Arbo2010}. Upon strong field ionization by a single-color light field at a central frequency of $2\omega$ the electron energy spectrum shows discrete energy peaks that are separated by $\Delta E=2 \hbar \omega$ ($T_{390}=\frac{2\pi}{2\omega}$, $2\omega=0.1168$\,a.u. corresponds to the frequency of a light field at a wavelength of 390\,nm). These discrete electron energies appear as concentric rings in the electron momentum distribution in the plane of polarization. 

\begin{figure}
\epsfig{file=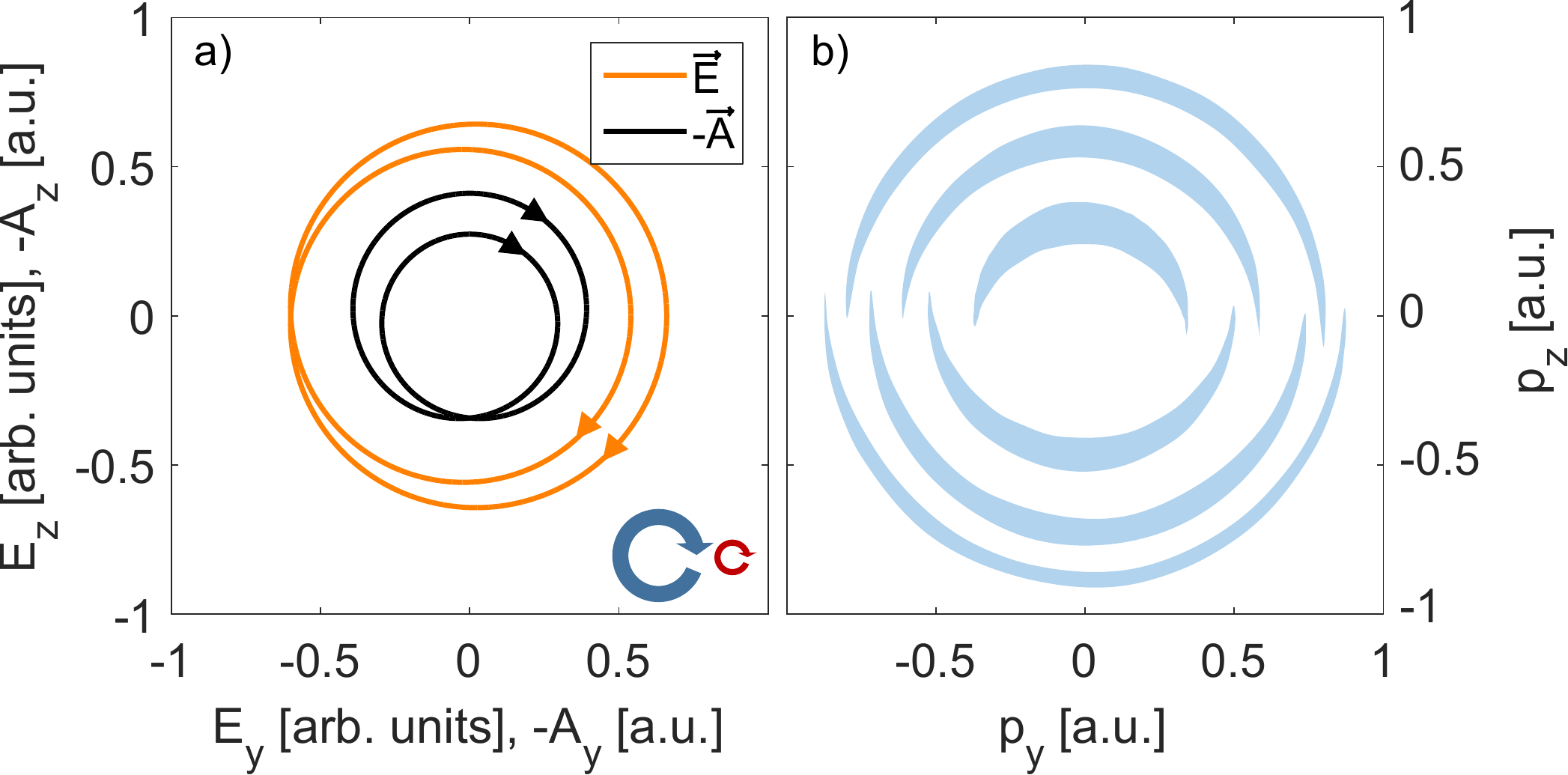, width=8.7cm}
\caption{(a) shows the combined electric field $\vec{E}$ and the negative vector potential $-\vec{A}$ of a co-rotating two-color (CoRTC) field. The helicities of the two colors and the temporal evolution of $\vec{E}$ and $-\vec{A}$ are indicated with arrows. (b) shows a sketch of the alternating half-ring (AHR) pattern in momentum space that is expected for a CoRTC field that is dominated by the light field at a central frequency of $2\omega$ (colored regions indicate high intensity in final momentum space).}
\label{fig_figure1label} 
\end{figure}

Using two-color fields that consist of a high-intensity light field at a central frequency of $2\omega$ and a low-intensity light field at a central frequency $\omega$ (typically the intensities differ by a factor of 100) leads to the appearance of sidebands between the energy peaks caused by the $2\omega$ field \cite{Zipp2014}. Light fields that consist of a fundamental and a second harmonic frequency that have the same helicity are referred to as co-rotating two-color (CoRTC) fields (see Fig. \ref{fig_figure1label}(a) for an example). In this case, the intensity of these rings in the electron momentum distribution is modulated as a function of the angle in the plane of polarization \cite{Han2018}. This results in an alternating half-ring (AHR) pattern in momentum space as illustrated in Fig. \ref{fig_figure1label}(b). The origin of this pattern is a sub-cycle interference which has been observed experimentally and reproduced using saddle-point strong field approximation as well as by solving the time-dependent Schr\"odinger equation \cite{Han2018,Ge2019}. Recently, Feng et al. \cite{Feng2019} have succeeded to model the appearance of sidebands for linearly polarized light by considering semi-classical trajectories that have release times that differ exactly by $T_{390}$ which is one cycle of the $2\omega$ field.

Here, we build on the perspective of interference of wave packets in momentum space \cite{Huismans2011, Eckart2018SubCycle, ValenceElectronMotion2018} and refer to this as holographic angular streaking of electrons (HASE). It will be shown that - by using the framework of HASE - changes of the Wigner time-delay \cite{Wigner1955,Pazourek2015}) become accessible also in the multi-photon and tunneling regime by measuring final electron momentum distributions \cite{Eppink1997,jagutzki2002multiple,ullrich2003recoil}.

The paper is organized as follows. In section 2, we present a trajectory-based semi-classical model (HASE model) which explains the alternating half-ring (AHR) pattern as interference between four different trajectories. In this model, we introduce one free parameter, external to the model, which is the initial phase of the trajectories. Section 3 shows the results of our HASE model for the case that this initial phase is set to zero. Section 4 shows how the initial phase modifies the AHR pattern. Section 5 illustrates how the initial phase is connected to a position offset of the initial wave packet, that is modeled by the trajectories.  Section 6 illustrates how the initial phase is related to the Wigner time delay, and section 7 provides a recipe, explaining how to use our formalism to obtain the Wigner time delay in strong-field ionization from the observable AHR pattern. We conclude with a discussion of the influence of Coulomb interaction in section 8. The abbreviation ``a.u.'' is used to indicate atomic units throughout the manuscript.
\section{II. Theoretical Model}
\begin{figure}
\epsfig{file=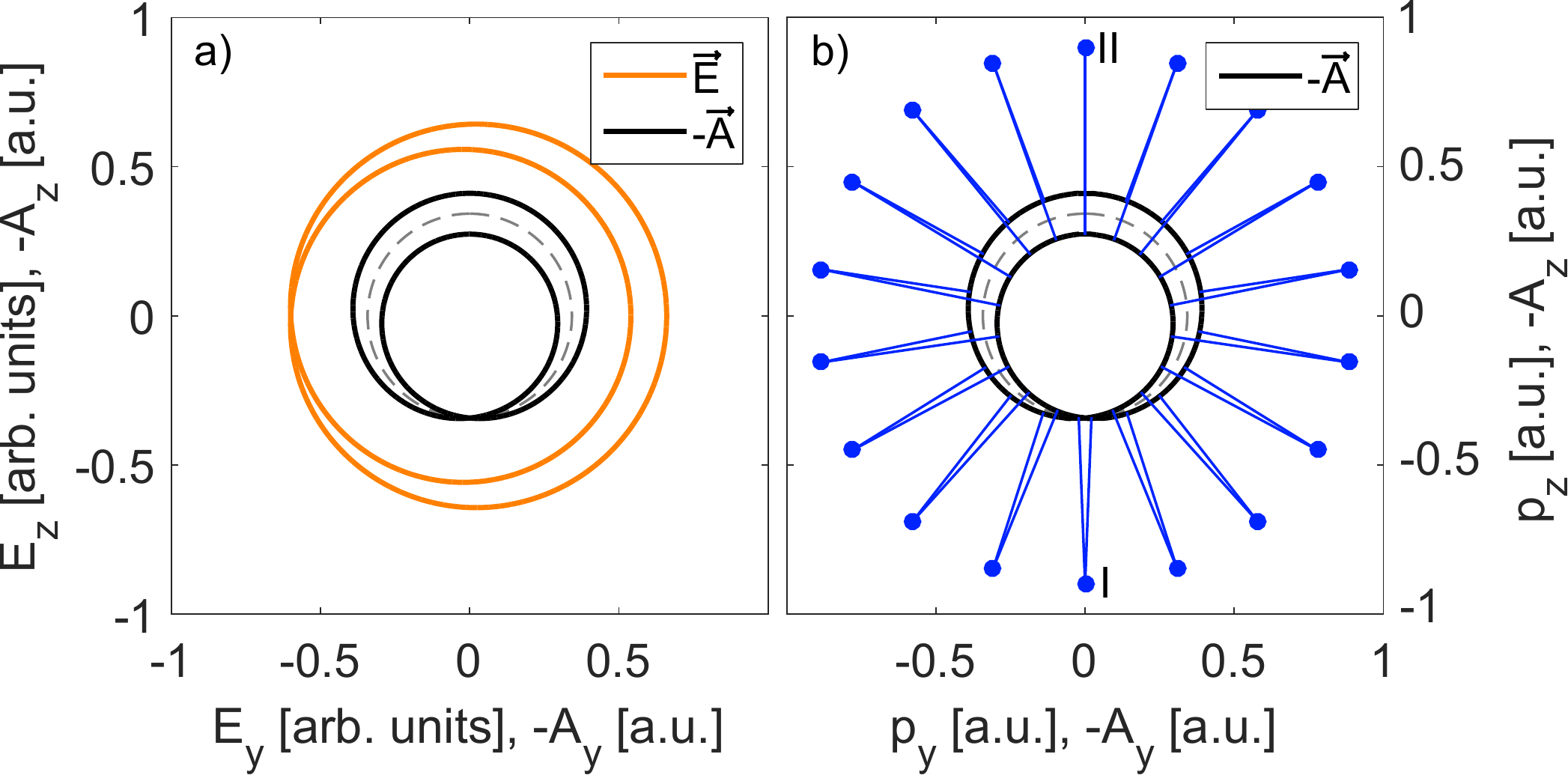, width=8.7cm}
\caption{(a) shows the combined electric field $\vec{E}(t)$ and the negative vector potential $-\vec{A}(t)$. The blue dots in (b) indicate possible final electron momenta $\vec{p}_f$. Each indicated value of $\vec{p}_f$ is connected by two blue lines to the values $-\vec{A}(t_1)$ and $-\vec{A}(t_2)$. $t_1$ and $t_2$ are the two release times that have been obtained from solving Eq. \ref{lab_initial} numerically. The vectors of the blue lines reflect the corresponding initial momenta $\vec{p}_{i,1}$ and $\vec{p}_{i,2}$.}
\label{fig_figure2label} 
\end{figure}

The time-dependent electric field, $\vec{E}(t)$, used throughout this work is illustrated in Fig. \ref{fig_figure2label}(a), together with the corresponding negative vector potential $-\vec{A}(t)$ which is given in Eq. \ref{eqvecpot}. $\vec{E}(t)$ and $\vec{A}(t)$ are linked by $\vec{E}=-\frac{\mathrm{d}\vec{A}}{\mathrm{dt}}$.
\begin{equation}
\begin{aligned}
\vec{A}(t)&=-\begin{bmatrix}
           0 \\
           \frac{E_{780}}{\omega} \sin(\omega t)+\frac{E_{390}}{2\omega} \sin(2\omega t) \\
           \frac{E_{780}}{\omega} \cos(\omega t)+\frac{E_{390}}{2\omega} \cos(2\omega t)
         \end{bmatrix}
\end{aligned}
\label{eqvecpot}
\end{equation}
Here, $t$ is the time and $\omega$ is the angular frequency of light at a wavelength of 780\,nm (the field amplitudes are $E_{390}=0.04$\,a.u. and  $E_{780}=0.004$\,a.u. for the two-color field and $E_{390}=0.04$\,a.u. and  $E_{780}=0.0$\,a.u. for the single-color field throughout this work).

In sections 1-7 of this paper, a simplified two-step model is used (HASE model): the electron is set free by tunnel ionization and then accelerated by the laser field. For the propagation after tunneling the Coulomb interaction is neglected.  Thus, the final electron momentum $\vec{p}_f$ is the sum of the negative vector potential at the time the electron tunnels $-\vec{A}(t)$ and the initial momentum $\vec{p}_i$ after tunneling. The initial momentum $\vec{p}_i$ must be perpendicular to the laser electric field $\vec{E}(t)$ at the instance of tunneling. (The initial momentum component perpendicular to the yz-plane ($p_x$) is set to zero in our HASE model.) This leads to Eq. \ref{lab_initial}. If one considers only release times $t$ within a single light cycle ($0\,as <t<T_{780}$, with $T_{780}=2\pi/\omega\approx 2602$\,as), then there are two release times $t_1 \neq t_2$ that lead to the same final electron momentum $\vec{p}_f$. Each release time $t_n$ must fulfill Eq. \ref{lab_initial} to ensure that the initial momentum along the tunneling direction (which is anti-parallel to the electric field at the instance of tunneling) is zero (trajectory number $n \in {1,2,3,4}$). 
\begin{equation}
\begin{aligned}
\vec{p}_f&=-\vec{A}(t_n)+\vec{p}_{i,n}\\
&=-\vec{A}(t_n)+\frac{p_n}{|\vec{E}(t_n)|}\begin{pmatrix}0\\E_z(t_n) \\ E_y(t_n) \end{pmatrix}
\end{aligned}
\label{lab_initial}
\end{equation}
Here, $p_n$ defines the absolute value and the sign of the initial momentum after tunneling. The sign of $p_n$ is defined such that a positive value of $p_n$ corresponds to a case in which $-\vec{A}(t_n)$ and $\vec{p}_{i,n}$ are parallel and therefore $|\vec{p}_f|>|\vec{A}(t_n)|$ for the field geometry shown in Fig. \ref{fig_figure2label}(a). In full analogy, negative values of $p_n$ lead to $|\vec{p}_f|<|\vec{A}(t_n)|$.

Solving Eq. \ref{lab_initial} numerically leads to two possible initial momenta ($\vec{p}_{i,1}$ and $\vec{p}_{i,2}$) for every final electron momentum $\vec{p}_f$. This gives rise to a two path interference. The two possibles pathways to the same final electron momentum are illustrated in Fig. \ref{fig_figure2label}(b): For several final electron momenta the two possible vector potentials that lead to these final electron momenta are indicated by connecting the respective $-\vec{A}(t_1)$ and $-\vec{A}(t_2)$ (using blue lines) with the final electron momenta (indicated by blue dots). Interestingly, the lines for the dot that is labeled with I in Fig. \ref{fig_figure2label}(b) have a finite intermediate angle. Closer inspection reveals that this is the case for all pairs of lines except for the dot that is labeled with II in Fig. \ref{fig_figure2label}(b).

To be able to model intra- and inter-cycle interference on the same footing using a semi-classical model we consider four wave packets (release times $t_1$, $t_2$, $t_3=t_1+T_{780}$, $t_4=t_2+T_{780}$, $\vec{p}_{i,1}=\vec{p}_{i,3}$ and $\vec{p}_{i,2}=\vec{p}_{i,4}$). The phases of the four semi-classical trajectories (trajectory number $n \in {1,2,3,4}$) at the time $t_f=t_4$ are modeled by using Eq. \ref{phasemodeling}. See e.g. Ref. \cite{Shilovski2016,Ni2016,Ni2018_theo,Shilovski2019} for an overview regarding semi-classical trajectories.
\begin{equation}
\begin{aligned}
\phi_n(\vec{p}_f,t_f)&=\frac{I_p t_n}{\hbar}-\phi_{\mathrm{prop}} (\vec{p}_f,t_n,t_f)+\phi_{\mathrm{off}}\\
\end{aligned}
\label{phasemodeling}
\end{equation}

\begin{figure*}
\epsfig{file=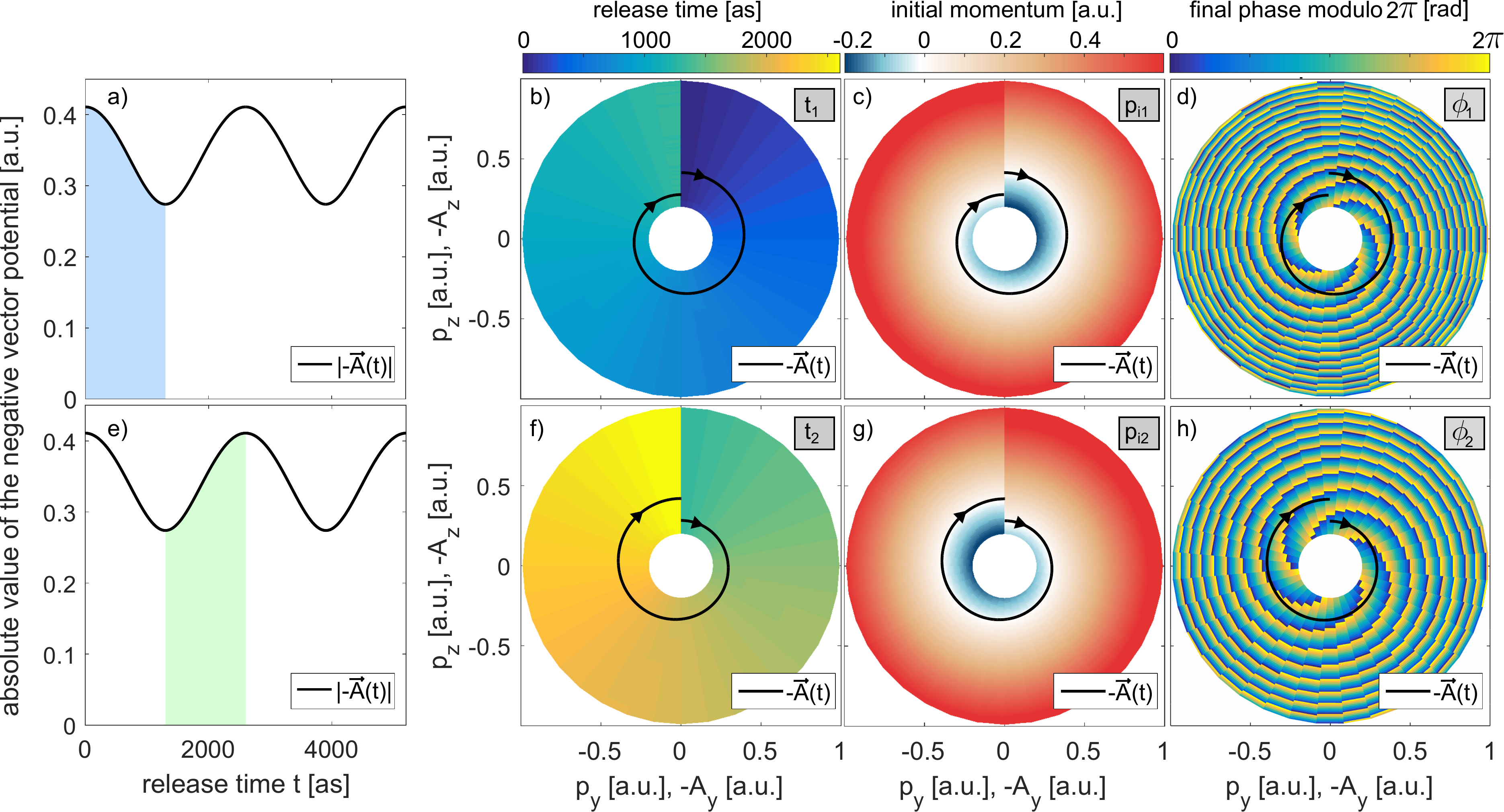, width=17.9cm} 
\caption{(a) illustrates the absolute value of the negative vector potential $|-\vec{A}(t)|$ and highlights the first half-cycle ($0\mathrm{\,as}<t\leq T_{390}$) of the two-cycle light field in blue. $|-\vec{A}(t)|$ is shown as a black line in all panels. (b) [(c)] shows the electron release time [initial momentum] as a function of the final electron momentum. (d) depicts the final phase modulo $2\pi$ as a function of the final electron momentum. (e)-(f) show the same as (a)-(d) but using the second half-cycle ($T_{390}<t \leq 2 T_{390}$). The black arrows in (b)-(d) and (f)-(h) indicate the temporal evolution of $-\vec{A}(t)$.}
\label{fig_figure3label} 
\end{figure*}

Here,  $I_p$ denotes the ionization potential ($I_p=15.76$\,eV is used, which is the ionization potential of argon) and $\phi_{\mathrm{off}}$ is an offset phase (for the sake of simplicity this offset phase can considered to be zero until the discussion of Fig. \ref{fig_linearphasemomdistribution}). $I_p t_n$ models the phase evolution of the electron in its bound state. $\phi_{\mathrm{prop}} (\vec{p}_f,t_n,t_f)$ describes the change of the electron's phase after tunneling starting from the release time $t_n$ until the final time $t_f$. At the time $t_f$ the electron possesses the final momentum $\vec{p}_f$. Throughout this paper the final time is set to $t_f=t_4$. For every considered trajectroy ($n \in {1,2,3,4}$), the change in phase after tunneling is described by the integral of the electron's energy over time \cite{Shilovski2016}:
\begin{equation}
\begin{aligned}
\phi_{\mathrm{prop}}& (t_n,t_f,\vec{p}_f) = \frac{1}{\hbar}\int_{t_n}^{t_f} \frac{p_y^2(t)+p_z^2(t)}{2 m_e}dt\\
&=\frac{1}{\hbar}\int_{t_n}^{t_f} \frac{(A_y(t)+p_{yf})^2+(A_z(t)+p_{zf})^2}{2 m_e}dt
\end{aligned}
\label{lab1}
\end{equation}
Here, $m_e=1$\,a.u. is the electron's mass and $\hbar=1$\,a.u. is the reduced Planck constant.
For Eq. \ref{lab1} it is used that the instantaneous momentum can be expressed by the difference of the instantaneous vector potential $\vec{A}(t)$ and the final momentum $\vec{p}_f$. Thus, the semi-classically modeled wave function at a given final electron momentum $\vec{p}_f$ is given by:
\begin{equation}
\begin{aligned}
\Psi(\vec{p}_f)= \sum_{n=1}^4 B(p_n) \exp(i\phi_n(\vec{p}_f,t_f))
\end{aligned}
\label{psisuqared}
\end{equation}
Here, $B(p_n)$ is the amplitude which is given by the square root of the existence probability of the respective trajectory (for the sake of simplicity the amplitude $B(p_n)$ can considered to be one until the discussion of Fig. \ref{fig_linearphasemomdistribution}).
Finally, the experimentally accessible intensity in final electron momentum space is modeled by $|\Psi(\vec{p}_f)|^2$.

\begin{figure*}
\epsfig{file=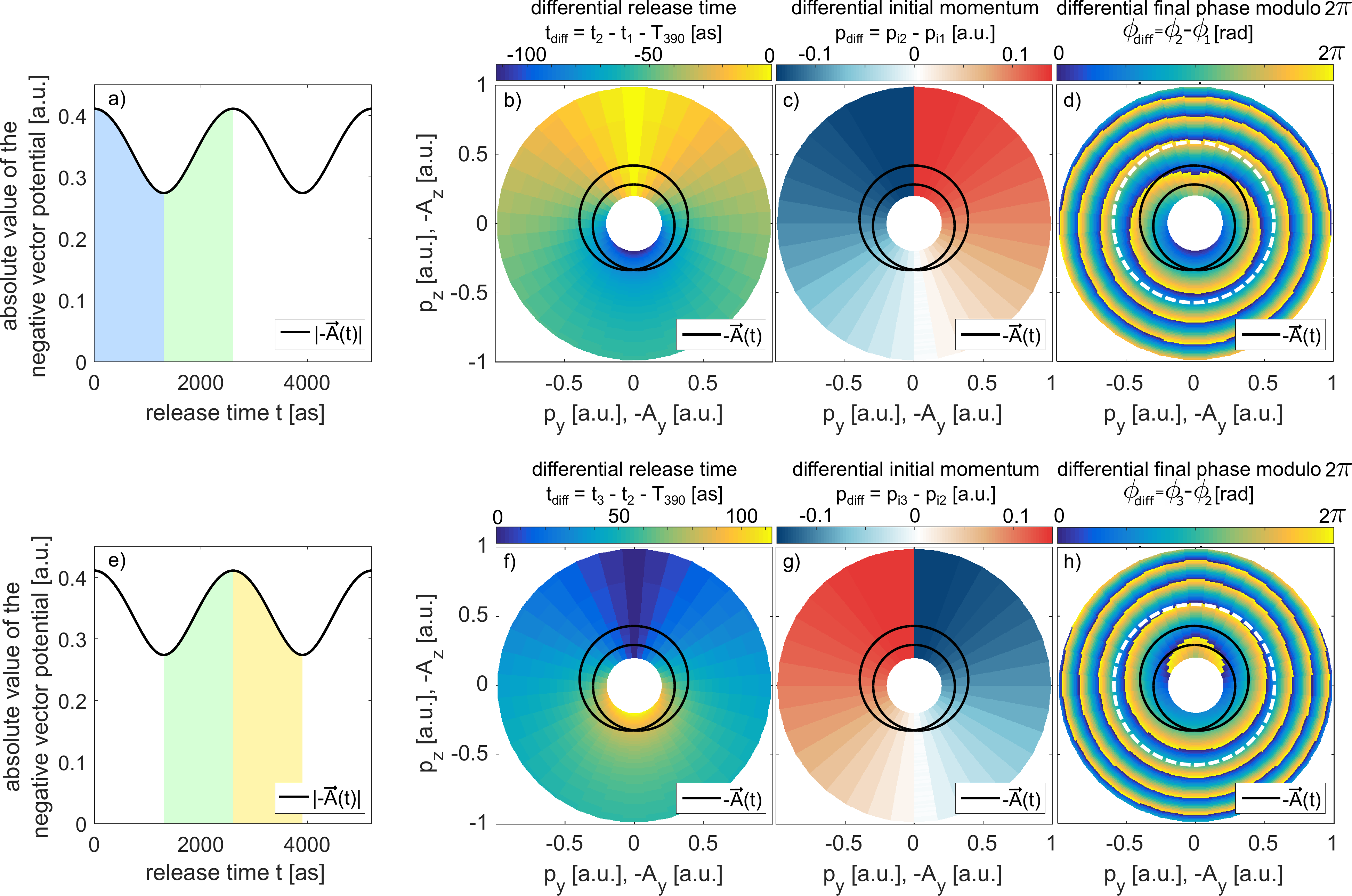, width=17.9cm} 
\caption{(a) the relevant ionization times for the first row are highlighted. The first (second) half-cycle is highlighted in blue (green). (b) shows the differential release time of Fig. \ref{fig_figure3label}(b) and \ref{fig_figure3label}(f). (c) shows the differential initial momentum of Fig. \ref{fig_figure3label}(c) and \ref{fig_figure3label}(g). (d) shows the differential final phase comparing Fig. \ref{fig_figure3label}(d) and \ref{fig_figure3label}(h). (e)-(h) are analogous to (a)-(d) but compare the third and the second half-cycle instead of the second and the first half-cycle. The dashed white lines in (d) and (h) are circles to guide the eye.}
\label{fig_figure4label} 
\end{figure*}

\section{III. Numerical Examples without Offset Phase}
In the following, the newly introduced theoretical HASE model is used to produce numerical results. In all examples two optical cycles of the two-color field are considered. The absolute value of the negative vector potential $|-\vec{A}(t)|$ is shown as a black line in Fig. \ref{fig_figure3label}(a) and the first half-cycle ($0\mathrm{\,as}<t_1\leq T_{390}$) is highlighted in blue. For radial momenta $p_r=\sqrt{p_y^2+p_z^2} \in [0.2,1]$ Eq. \ref{lab_initial} is solved numerically. The release time $t_1$ is depicted as a function of the final electron momentum in Fig. \ref{fig_figure3label}(b). As expected, the release time increases monotonically with the angle in the plane of polarization \cite{Eckle2008}. Fig. \ref{fig_figure3label}(c) shows the initial momentum $\vec{p}_{i,1}$ as a function of the final electron momentum. Using Eq. \ref{phasemodeling} with an offset phase of zero ($\phi_{\mathrm{off}}=0$\,rad) allows for the calculation of the phase as function of the final electron momentum (see Fig. \ref{fig_figure3label}(d)). 

Fig. \ref{fig_figure3label}(e)-(h) are generated in analogy to Fig. \ref{fig_figure3label}(a)-(d) but here the release time is restricted to $T_{390}<t_2\leq 2T_{390}$. In Fig. \ref{fig_figure3label}(c) and (g) it can be nicely seen that the value of the initial momentum is zero at final momenta that coincide with the negative vector potential and increases (decreases) for increasing (lower) radial momenta.

In Fig. \ref{fig_figure4label}(a)-(d) the differences of the first and the second row in Fig. \ref{fig_figure3label} are presented. The ionization times that are compared are highlighted in Fig. \ref{fig_figure4label}(a). The differences in release time $t_{\mathrm{diff}}=t_2-t_1-T_{390}$ are shown in Fig. \ref{fig_figure4label}(b) as a function of the final momentum. Strikingly, the difference is not zero but values of more than 100\,attoseconds are found. The differences in the initial momentum $p_{\mathrm{diff}}=p_{i,2}-p_{i,1}$ are shown in Fig. \ref{fig_figure4label}(c) as a function of final momentum and are as expected from the shape of the vector potential. The differences in final phase $\Phi_{\mathrm{diff}}=\Phi_2-\Phi_1$ are presented in Fig. \ref{fig_figure4label}(d) and show a slightly distorted circular symmetry (for a single-color laser field a perfect circular symmetric distribution would be observed). Fig. \ref{fig_figure4label}(e)-\ref{fig_figure4label}(h) are analogous to Fig. \ref{fig_figure4label}(a)-\ref{fig_figure4label}(d) but compare the third and the second half-cycle instead of the second and the first half-cycle (as visualized in Fig. \ref{fig_figure4label}(e)). Because of the definitions $t_3=t_1+T_{780}$ and $p_{i,3}$=$p_{i,1}$ the results in Fig. \ref{fig_figure4label}(f) and \ref{fig_figure4label}(g) are the same as Fig. \ref{fig_figure4label}(b) and \ref{fig_figure4label}(c) but with opposite sign. Fig. \ref{fig_figure4label}(d) and \ref{fig_figure4label}(h) are not trivially linked because the final electron phase has to be evaluated using Eq. \ref{phasemodeling}.

\begin{figure}
\epsfig{file=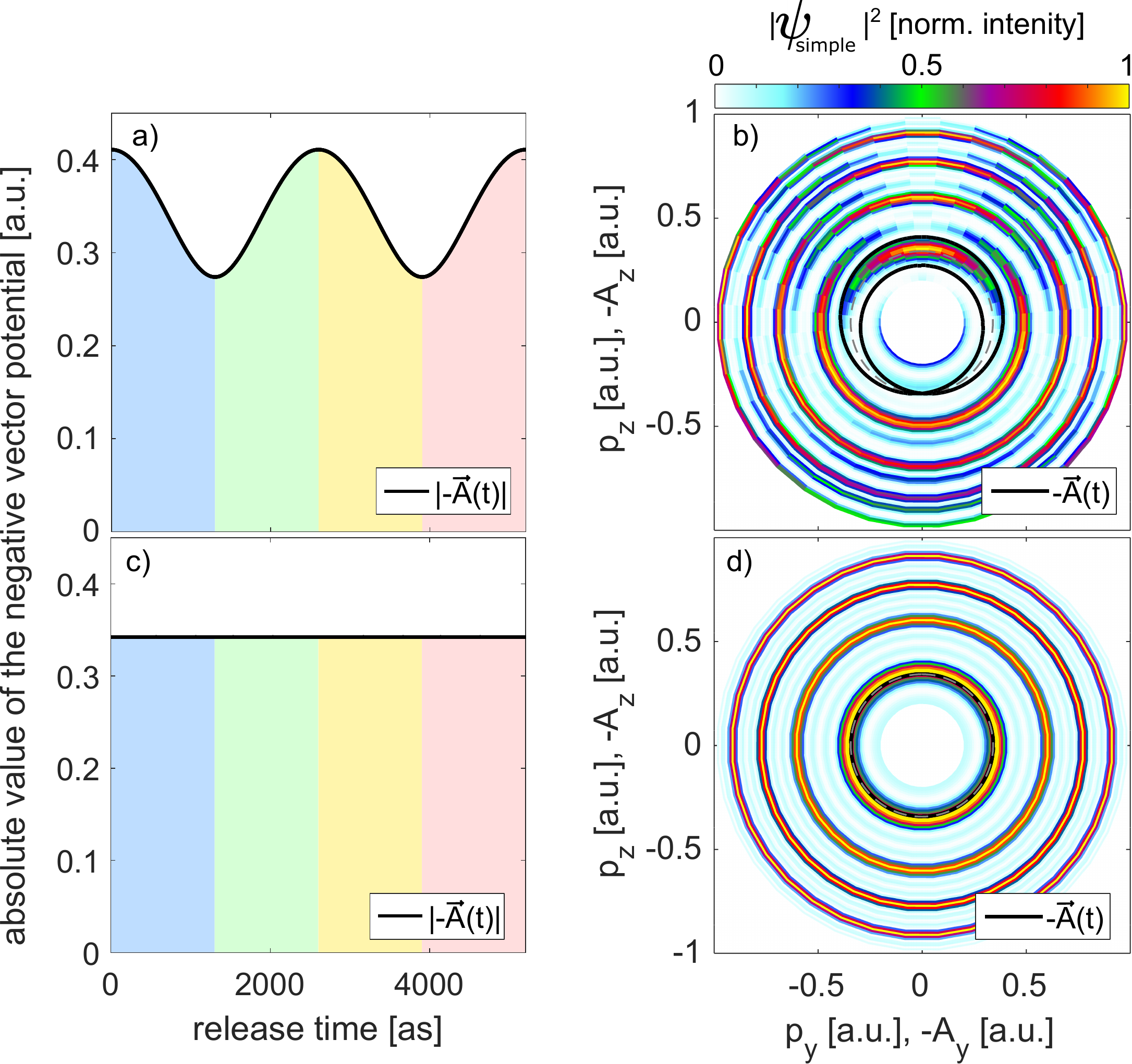, width=8.7cm} 
\caption{(a) illustrates that electron release times from all half-cycles are considered. (b) shows the electron momentum distribution $|\Psi_{\mathrm{simple}}|^2$ with the expected alternating half-rings. (c) and (d) show the same as (a) and (b) but using a single-color field with $E_{390}=0.04$\,a.u. The black line represents the negative vector potential and the gray dashed line guides the eye and is the same in (b) and (d).}
\label{fig_singlevstwocolor} 
\end{figure}

The experimentally accessible quantity is $|\Psi(\vec{p}_f)|^2$. Assuming that all four trajectories have the same amplitude ($B(p_n)=1$) and using an offset phase of zero ($\phi_{\mathrm{off}}=0$\,rad) Eq. \ref{psisuqared} can be evaluated for each final electron momentum $\vec{p}_f$. Each trajectory is released within one of the four half-cycles of the laser field (as illustrated in Fig. \ref{fig_singlevstwocolor}(a)). The result is presented in Fig. \ref{fig_singlevstwocolor}(b) as  $|\Psi_{\mathrm{simple}}|^2$ and the expected alternating half-rings in final electron momentum space are reproduced. For comparison a single-color field at 390\,nm with $E_{390}=0.04$\,a.u. is evaluated that leads to the well-known ATI structure in final electron momentum space without any angular modulations or sidebands (see Fig. \ref{fig_singlevstwocolor}(d)). 

\section{IV. Numerical Examples with Offset Phase}
\begin{figure}
\epsfig{file=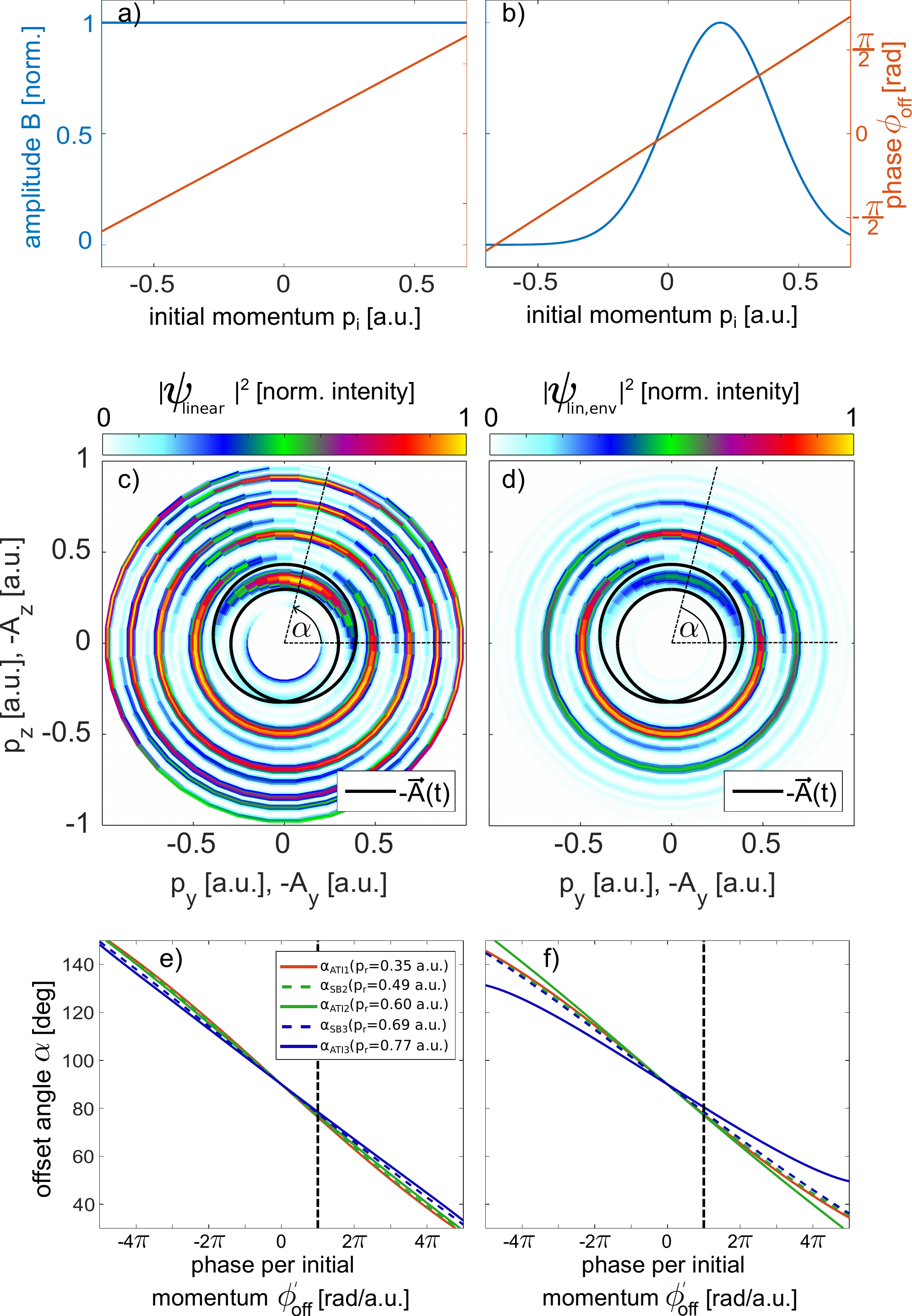, width=8.7cm} 
\caption{(a) [b] shows an initial momentum distribution with a constant [non-constant] amplitude $B(p_i)$ and a linear phase. (c) [(d)] shows the expected intensity in final momentum space $|\Psi_{\mathrm{linear}}|^2$ [$|\Psi_{\mathrm{lin,env}}|^2$] using the initial momentum distribution shown in (a) [(b)]. (e) [(f)] shows the offset angles $\alpha$ that have been extracted from electron momentum distributions as in (c) [(d)]. The offset angles of the ATI peaks are determined as indicated in (c) and (d). The offset angles of the sideband peaks are determined in full analogy but subtracting $\pi$ to take the alternating pattern of ATI peaks and sidebands into account (see section 7 for details).}
\label{fig_linearphasemomdistribution} 
\end{figure}

In this section the influence of a non-zero offset phase $\phi_{\mathrm{off}}$ is investigated. We use an offset phase that depends linearly on the initial momentum $p_{i}$ by setting $\phi_{\mathrm{off}}(p_{i})=\kappa \frac{p_{i}}{a.u.}$ (for a real valued $\kappa$). Fig. \ref{fig_linearphasemomdistribution}(a) shows the phase of such such an initial momentum distribution with $\kappa=\pi$  that has a constant amplitude of $B(p_n)=1$. Evaluating $|\Psi(\vec{p}_f)|^2$ for these parameters leads to $|\Psi_{\mathrm{linear}}|^2$ which is shown in Fig. \ref{fig_linearphasemomdistribution}(c). Strikingly, a rotation of the electron momentum distribution with respect to Fig. \ref{fig_singlevstwocolor}(b) can be seen. Fig. \ref{fig_linearphasemomdistribution}(e) quantifies the rotations of the ATI peaks and the sidebands in more detail by plotting the offset angle $\alpha$ as a function of $\phi^{\prime}_{\mathrm{off}}=\frac{\partial \phi_{\mathrm{off}}}{\partial p_{i}}$ (separately for every energy peak). The value of the offset angle $\alpha$ is retrieved using the following procedure: the angular distribution for every energy peak is analyzed separately by performing a Fourier transformation to extract the offset angle $\alpha$ from the phase of the lowest frequency component in Fourier space (not the DC component, see section 7 for details and note that $\alpha$ is defined as indicated in Fig. \ref{fig_linearphasemomdistribution}(c) and \ref{fig_linearphasemomdistribution}(d)).

In general, $\phi^{\prime}_{\mathrm{off}} (p_i)$ does not have to be constant but can vary with $p_i$. Using the driving field shown in Fig. \ref{fig_figure1label}, for any absolute value of the final electron momentum $|\vec{p}_f|$ the two relevant initial momenta ($p_{i1}$ and $p_{i2}$), that lead to this final momentum, differ by less than $0.15$\,a.u. (see Fig. \ref{fig_figure4label}(c) and \ref{fig_figure4label}(g)). Hence, the offset angle $\alpha$ can be used to infer the value of $\phi^{\prime}_{\mathrm{off}}(p_{i})$ but represents not the exact derivative but approximates the deviate in in interval with a length of upto $\Delta p_{i}=0.15$\,a.u. (In principle the value of $\Delta p_{i}$ could be further reduced by decreasing the intensity of the light field at at central frequency of $\omega$, because this also reduces the difference of the minimal and the maximal value of the absolute value of the negative vector potential $|-\vec{A}|$.)

Inspecting Fig. \ref{fig_linearphasemomdistribution}(a) the choice of $B(p_{i})=1$ appears to be unrealistic. More realistic values of the amplitude distribution are shown in Fig. \ref{fig_linearphasemomdistribution}(b) using $B(p_{i})=\exp \left(\frac{\left(p_{i}-p_{\mathrm{0}}\right)^2}{2\sigma^2}\right)$. Here, $\sigma=0.2$\,a.u. accounts for the width of the initial momentum distribution after tunneling \cite{Arissian2010} and $p_{\mathrm{0}}=0.2$\,a.u. is chosen to model a typical non-adiabatic momentum offset \cite{Olga2011A,Olga2011B,Eckart2018_Offsets}. The result is shown as $|\Psi_{\mathrm{lin,env}}|^2$ in Fig. \ref{fig_linearphasemomdistribution}(d). As expected, this mainly affects the visibility of the inner and outer energy peaks (compare intensity envelopes of Fig. \ref{fig_linearphasemomdistribution}(c) and (d)). Interestingly, the rotation angles $\alpha$ are hardly affected (compare Fig. \ref{fig_linearphasemomdistribution}(e) and (f)). It can be concluded, that for typical \cite{Han2018,Ge2019,DanielArXiv2020,EckartArXivSideband} light intensities of CoRTC fields the derivative of the phase of the initial momentum distribution $\phi^{\prime}_{\mathrm{off}}(p_i)$ can be inferred from the (experimentally accessible) offset angles $\alpha$ in final momentum space.

For the conditions that are used throughout this paper, the Keldysh parameter, $\gamma=\frac{\omega_{\mathrm{eff}}}{e E_0}\sqrt{2 m_e I_p}$, is close to $\gamma=3$. Here, $\omega_{\mathrm{eff}}$ is the effective angular frequency that is close to the angular frequency of light at a wavelength of 390\,nm (see Ref. \cite{Eckart2018_Offsets} for details). This is the regime of non-adiabatic tunneling \cite{Misha2005,Olga2011A,Eckart2018_Offsets}. In our HASE model, the electronic wave packet after tunneling is described by an initial momentum dependent amplitude and an initial momentum dependent phase (see Fig. \ref{fig_linearphasemomdistribution}(a) and \ref{fig_linearphasemomdistribution}(b)). A typical non-adiabatic offset of the amplitudes in momentum space can be modeled as shown in Fig. \ref{fig_linearphasemomdistribution}(b). In our HASE model the phase of the initial wave packet in momentum space is defined by Eq. \ref{phasemodeling} (for $\phi_{\mathrm{off}}=0$\,rad this is equivalent to the typical assumption for adiabatic tunneling, see e.g. Ref. \cite{Shilovski2016}). It is important to realize that the phase of the wave packet might be affected by the non-adiabaticity of the tunneling process. However, $\phi_{\mathrm{off}}(p_i)$ is external to the HASE model which allows for the modeling of an arbitrary phase structure of the electron wave packet after tunneling and calculate the resulting interference pattern in the final electron momentum distribution. Setting $\phi^{\prime}_{\mathrm{off}}$ to certain values that are independent of $p_i$ and investigating $\alpha$ is just one possibility (as it has been done for Fig. \ref{fig_linearphasemomdistribution}). In turn, hypotheses about the electron wave packet's phase structure upon non-adiabatic tunneling might be tested using our HASE model or the SCTS model presented in section 8. We emphasize, that we do not claim, nor prove that our model is exact. We suggest viewing our theoretical model as a simple man's model to holographic angular streaking that should be benchmarked by comparison with experiments \cite{EckartArXivSideband,DanielArXiv2020} and future theoretical studies (also see section 8).

It should be noted that the sub-cycle dependence of the ionization rate \cite{Yudin2001A} and Coulomb interaction after tunneling \cite{torlina2015interpreting,Bray2018,Ni2018_theo} are not included in the HASE model. The sub-cycle dependence of the ionization rate is expected to lead to similar deviations as the envelope of the amplitudes of the initial momentum distribution (see Fig. \ref{fig_linearphasemomdistribution}(a) and \ref{fig_linearphasemomdistribution}(b)). In section 8 we present an SCTS model that includes Coulomb interaction after tunneling and compare the results with the results of the HASE model. A validation of the HASE model using a full quantum simulation (e.g. by using the time-dependent Schr\"odinger equation) has not yet been achieved and is beyond the scope of this paper.

\section{V. HASE and Offsets of the Bound Wave Function in Position Space}
As described above CoRTC fields allow one to obtain $\phi^{\prime}_{\mathrm{off}}=\frac{\partial \phi_{\mathrm{off}}}{\partial p_{\mathrm{i}}}$ from measured electron momentum distributions. The next step is to understand the \textit{meaning} of a phase gradient of the initial momentum distribution.

The only position-space-information that is explicitly included in the HASE model is that the initial momentum distribution is zero along the direction of the electric field at the instance of tunneling. However, there is more position-space-information included in the model because a linear phase in momentum space corresponds to a shift in position space (complex momentum space and complex position space are linked by Fourier transformation, also see Refs. \cite{ValenceElectronMotion2018,MaksimNatureCom} for similar approaches to measure position-information). Let the position dependent wave function $\Psi(x_i)$ be the Fourier transform of the momentum dependent wave function $\Psi(p_i)$. Then a real valued position offset $\Delta x$ leads to the shifted position dependent wave function $\bar{\Psi}(x_i)=\Psi (x_i+\Delta x)$. The Fourier transform leads to the wave function $\bar{\Psi} (p_i)$.
\begin{equation}
\begin{aligned}
\bar{\Psi} (p_i)=\exp \left( i \frac{p_i \Delta x}{\hbar} \right) \Psi (p_i)
\end{aligned}
\label{fftshift}
\end{equation}

\begin{figure}
\epsfig{file=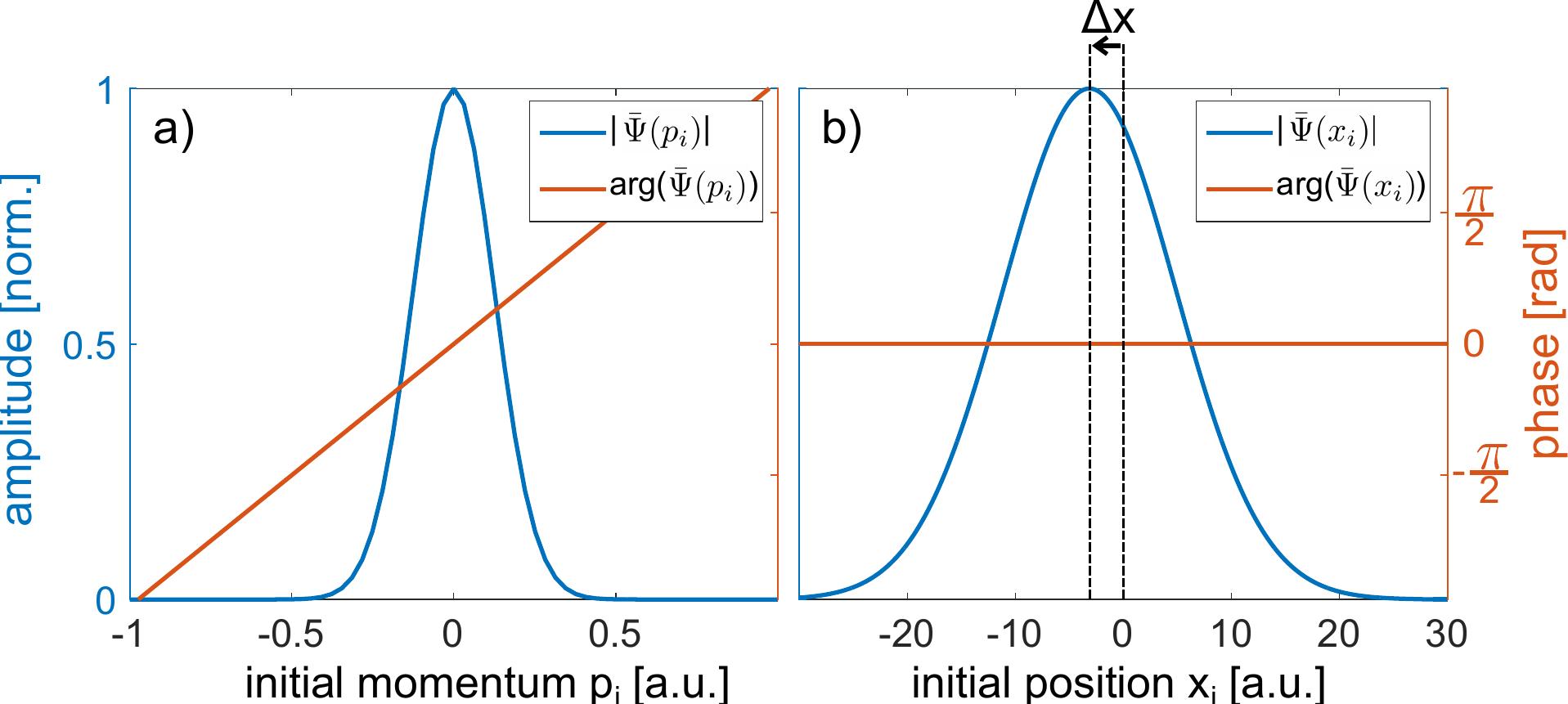, width=8.7cm} 
\caption{(a) shows a wave function for an initial momentum distribution with a Gaussian distribution of the amplitudes and a linear phase. (b) shows the Fourier transform of the distribution that is shown in (a). The black vertical lines guide the eye to emphasize the shift of the amplitudes in (b) that is proportional to the slope of the phase in (a) (see Eq. \ref{fftshift}).}
\label{fig_fftmompos} 
\end{figure}

Fig. \ref{fig_fftmompos} illustrates the relation that is described in Eq. \ref{fftshift} by showing a wave function for the initial momentum distribution with a Gaussian distribution of the amplitudes that is centered around zero initial momentum. For a constant phase the Fourier transform of a Gaussian distribution would be another Gaussian distribution that is centered at zero. The linear phase in Fig. \ref{fig_fftmompos}(a) (same phase dependence as in Fig. \ref{fig_linearphasemomdistribution}(a) and (b)) is reflected by an offset in position space by $\Delta x$ as it can be seen in Fig. \ref{fig_fftmompos}(b).

Identifying $p_i \Delta x/\hbar=\phi_{\mathrm{off}}$ allows to link $\Delta x=\phi^{\prime}_{\mathrm{off}}\hbar$ for the case of a linear phase of the initial momentum distribution as discussed regarding Fig. \ref{fig_linearphasemomdistribution}. Thus, it can be concluded that the value of $\Delta x=\phi^{\prime}_{\mathrm{off}}\hbar$ can be interpreted as a measure of the displacement of the wave packet after tunneling in position space for the case of a linear phase of the initial momentum distribution. (Positive values of $\phi^{\prime}_{\mathrm{off}}$ correspond to a displacement of the wave packet after tunneling in position space that is anti-parallel to $-\vec{A}(t)$ at the electron release time $t$.)

Fig. \ref{fig_fftmompos} suggests that the measurement of the phase gradient in momentum space can be used to probe the amplitudes of the initial state's wave function in position space. This is the position space analogue to the much used fact that tunneling acts like a filter on the initial bound state's wave function in momentum space \cite{Arissian2010,Olga2011A,Fechner2014}, which has been used to infer fingerprints of the bound state's wave function in momentum space from the measured final state's momentum distribution (e.g. in Refs. \cite{Meckel2008,EckartNatPhys2018}). Thus, HASE is a novel approach to access e.g. molecular structure and polarization states in position space.

\begin{figure}
\epsfig{file=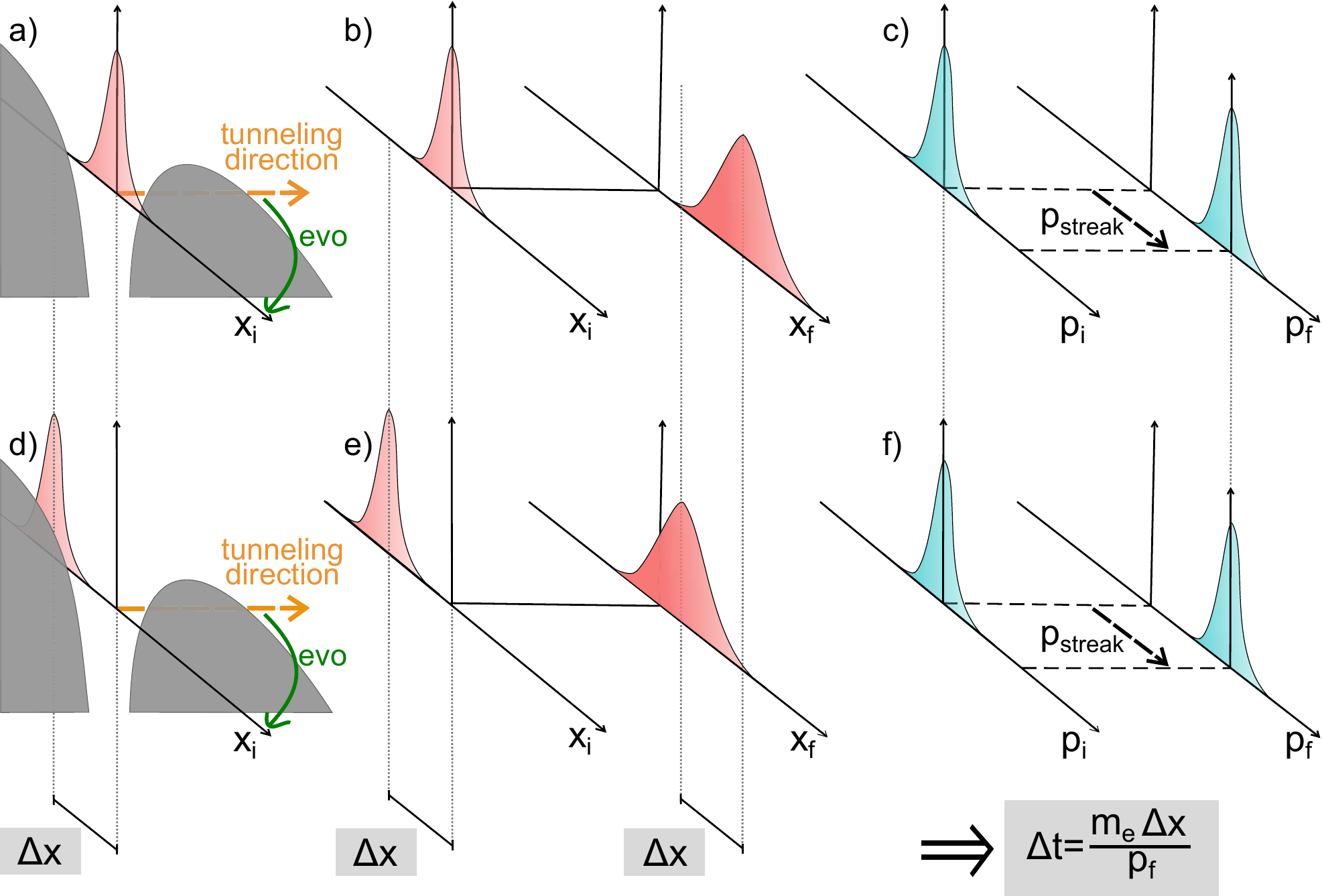, width=8.7cm}
\caption{(a) schematically illustrates an initial bound state in position space. $x_i$ is perpendicular to the tunneling direction. (b) depicts the position distribution along the same direction as in (a) directly after tunneling ($x_i$) and for the final time ($x_f$). At the final time $t_f$ the wave packet is broadened along $x_f$ due to dispersion of the continuum wave packet. Note, that the width of the distribution might change during tunneling which is neglected here for the sake of simplicity. (c) shows the initial momentum distribution along $p_i$ and the corresponding final momentum distribution along $p_f$ which only differ by the momentum that is due to the streaking of the laser $p_{\mathrm{streak}}$. (d)-(f) show the same as (a)-(c) with the only difference that the initial bound state in position space is displaced by $\Delta x$ as illustrated in (d). This shifts all positions in (e) by the value of $\Delta x$. The final momentum distributions in (c) and (f) are the same. The shift in position $\Delta x$ of a continuum wave packet with a given momentum $p_f$ allows to calculate the time delay of the scenario in (a)-(c) relative to the scenario in (d)-(f) using $\Delta t=\frac{m_e\Delta x}{p_f}$. The temporal evolution of the tunneling direction is indicated in (a) and (d) by the arrow labeled with ``evo''. }
\label{fig_overallscheme} 
\end{figure}

Looking at the entire ionization process as a \textit{black box} and only considering the continuum states allows for an illuminating insight: If Coulomb interaction after tunneling is neglected, any displacement of an initial state by a given vector leads to a displacement of the corresponding final state by the same vector. Let this displacement be anti-parallel to the direction of the streaking momentum $\vec{p}_{streak}=-\vec{A}(t)$, then, for a single-color circularly polarized field, a displacement of the initial position by $\Delta x$ is equivalent to a continuum wave packet that leaves the \textit{black box} with a time delay $\Delta t$. This can be described quantitatively using Eq. \ref{delaybypositionshift} and is illustrated in Fig. \ref{fig_overallscheme}.

\begin{equation}
\begin{aligned}
\Delta t&=\frac{m_e\Delta x}{p_f}\\
&=\frac{m_e \hbar}{p_f} \phi^{\prime}_{\mathrm{off}} (p_i)\\
&=\frac{m_e \hbar}{p_f} \phi^{\prime}_{\mathrm{off}} (p_f-p_{\mathrm{streak}})
\end{aligned}
\label{delaybypositionshift}
\end{equation}
Here, $m_e=1$\,a.u. is the electron's mass,  $\hbar=1$\,a.u. the reduced Planck constant and $p_{\mathrm{streak}}=|-\vec{A}|$. In the next section it will be shown that, within the HASE model, the time delay $\Delta t$ is the same as the Wigner time delay.

\section{VI. HASE and the Wigner Time Delay}
The generation of sidebands is closely related to reconstruction of attosecond harmonic beating by interference of two-photon transitions (RABBITT) \cite{vos2018orientation,Paul1689,Muller2002}. RABBITT can be used to access the Wigner time-delay \cite{Wigner1955}. By definition, RABBITT only treats two-photon transitions. In this section it is explained how HASE can be used to access changes of the Wigner time delay for multi-photon ionization and tunnel ionization.

So far it has been shown that HASE is sensitive to the slope of the phase of the initial momentum distribution, $\phi^{\prime}_{\mathrm{off}}$, and that $\phi^{\prime}_{\mathrm{off}}$ is linked to offsets of the initial position distribution. In the following the question how $\phi^{\prime}_{\mathrm{off}}$ affects the phase of the final, semi-classically modeled electronic wave function $\Psi (\vec{p}_f)$ is investigated. To answer this, we assume that the complex valued wave function of the initial momentum distribution does not change if the weak field at the central frequency $\omega$ is switched off.

The simple case of a single-color circularly polarized light field with $E_{390}=0.04$\,a.u. is considered and the derivative of the initial phase  $\phi^{\prime}_{\mathrm{off}}(p_i)$ is assumed to be known. In this case, the final electron momentum distribution is independent of the angle in the plane of polarization (see Fig. \ref{fig_singlevstwocolor}(d)). We now make use of the fact that changes in $\phi^{\prime}_{\mathrm{off}}(p_i)$ do not affect the trajectory or the probability $|\Psi (\vec{p}_f)|^2$ but only influence the final phase $\arg (\Psi (\vec{p}_f))$. 

Initial and final momenta are unambiguously linked by $\vec{p}_f=-\vec{A}(t)+\vec{p}_{i}$ (see Eq. \ref{lab_initial}). Because of the symmetry of circularly polarized single-color light fields, a phase gradient at a given value of $p_i$ directly allows to quantify the phase change at all final momenta with $|\vec{p}_{f}|=|\vec{A}(t)|+p_i$. This implies that the phase gradient of the final electron momentum distribution is known at the corresponding energy $E$. This energy dependent phase of the semi-classically modeled wave function can be expressed using the concept of the Wigner time delay $\tau_W$. Throughout this paper, the Wigner time delay $\tau_W$ is used as defined in Eq. \ref{Wignertimedefinition} (see Refs. \cite{Carvalho2002, Ivanov2013, cirelli2015energy,vos2018orientation} for similar interpretations of the Wigner time that are closely related to the group delay \cite{Ivanov2013, cirelli2015energy}).
\begin{equation}
\tau_W=\hbar \frac{\partial \arg (\Psi)}{\partial E}
\label{Wignertimedefinition}
\end{equation}

Using the trajectory-based HASE model and considering a single-color circularly polarized light field (as for Fig. \ref{fig_singlevstwocolor}(c) and (d)), the initial momentum is unambiguously linked with the final momentum. One can compare two scenarios: The first scenario uses $\phi^{\prime}_{\mathrm{off}}= 0$\,rad/a.u. leading to the semi-classically modeled wave function $\Psi_{\mathrm{simple}}(\vec{p}_f)$ at a given time $t_f$. In the second scenario an arbitrary phase of the initial momentum distribution $\phi^{\prime}_{\mathrm{off}}$ is used (the amplitudes $B(p_i)$ can be shown to be irrelevant regarding $\tau_W$ for a single-color circularly polarized light field) leading to the semi-classically modeled wave function $\Psi_{\mathrm{delayed}}(\vec{p}_f)$ at the same time $t_f$. The two semi-classically modeled wave functions are related by:
\begin{equation}
\Psi_{\mathrm{delayed}}(\vec{p}_f)=\Psi_{\mathrm{simple}}(\vec{p}_f) \exp (i \phi_\mathrm{off}(p_i))
\end{equation}

As a result, the change of the Wigner time delay due to the phase of the initial momentum distribution is given by:
\begin{equation}
\begin{aligned}
\Delta \tau_W&=\hbar \left( \frac{\partial \arg (\Psi_{\mathrm{delayed}} (\vec{p}_f))}{\partial E}-\frac{\partial \arg (\Psi_{\mathrm{simple}}(\vec{p}_f))}{\partial E}\right)\\
&=\hbar \frac{\partial \arg (\exp (i \phi_\mathrm{off}(p_i)))}{\partial E}\\
&=\hbar \frac{\partial \phi_\mathrm{off}(p_i)}{\partial E}
\end{aligned}
\end{equation}

Substituting energy with momentum leads to:
\begin{equation}
\begin{aligned}
\Delta \tau_W&=\hbar \frac{m_e}{p_f}\frac{\partial \phi_\mathrm{off}(p_i)}{\partial p_i}\\
&=\hbar \frac{m_e}{p_f}\phi_\mathrm{off}^{\prime}(p_i)\\
&=\hbar \frac{m_e}{p_f}\phi_\mathrm{off}^{\prime}(p_f-p_{\mathrm{streak}})
\end{aligned}
\label{finalWT}
\end{equation}
The result in Eq. \ref{finalWT} is equivalent to the previously obtained expression for the delay time $\Delta t$ (see Eq. \ref{delaybypositionshift}). Thus, we find that $\Delta t$ and  $\Delta\tau_W$ are equivalent within the HASE model. This result allows one to gain very fundamental insight: within the HASE model, the Wigner time delay for strong field ionization is related to an intuitive shift of the wave function in position space at the tunnel exit $\Delta x$ and might also be related to the bound wave function in position space.
We emphasize that the equivalence of $\Delta t$ and $\Delta\tau_W$ is not a result of a full quantum treatment and only a semi-classical result that is based on the HASE model (which neglects Coulomb interaction after tunneling). For realistic potential landscapes $\Delta t$ can deviate from $\Delta\tau_W$. A first approach to study such a scenario is presented in section 8.

\section{VII. Recipe for the Experimental Access to changes of the Wigner Time Delay in Strong Field Ionization}
The first step to access the Wigner time delay in strong field ionization within the framework of HASE is to conduct an experiment using a CoRTC field. High ratios of $E_{390}/E_{780}$ decrease the visibilty of the sidebands \cite{Feng2019} but also reduce the difference of the minimal and the maximal value of the absolute value of the negative vector potential $|\vec{A}(t)|$. As described above, higher values of $E_{390}/E_{780}$ lead to a more accurate mapping of the offset angle, $\alpha$, to the phase gradient of the initial momentum distribution after tunneling, $\phi_\mathrm{off}^{\prime}$. Typical values of $E_{390}/E_{780}$ are close to $10$ (see e.g. Refs. \cite{Han2018,Ge2019,DanielArXiv2020,EckartArXivSideband}). The amplitude of $E_{390}$ should be chosen such that the ionization channel that is investigated is not saturated. $E_{390}=0.04$\,a.u. is a good choice for the single ionization of argon \cite{Eckart2018_Offsets}.

Second, the offset angles $\alpha$ have to be retrieved from the measured final electron momentum distribution for every energy peak. (In principle, the result of a theoretical calculation that solves the time-dependent Schr\"odinger equation on a grid could be used as well. In this case the absolute square of the final electronic wave function in momentum space should be analyzed.) The final electron momentum distribution is considered to be in polar coordinates with $p_x$, $p_r=\sqrt{p_y^2+p_z^2}$ and $\phi_{\mathrm{polar}}$. Here, $\phi_{polar}$ is the angle in the yz-plane (defined in the same way as $\alpha$ in Fig. \ref{fig_linearphasemomdistribution}(c) and \ref{fig_linearphasemomdistribution}(d)). In a next step, the angular distribution for a given energy peak is analyzed. Let this angular distribution be represented by a vector $\vec{X}$ with N entries. The discrete Fourier transform $\vec{Y}$ is given by (where $k\in \{1, 2, ..., N\}$):
\begin{equation}
\label{eq1}
Y(k)=\sum_{j=1}^{N} X(j)  \exp \left( \frac{ (-2 \pi i) (j-1) (k-1)}{N} \right) 
\end{equation}
Now, the offset angle for the selected energy peak is given by $\alpha=-\arg\left( Y(2)\right)$ for the sidebands (energy peaks that vanish if $E_{780}$ is set to zero) and $\alpha=-\arg\left( Y(2)\right)-180\mathrm{\,deg}$ for the ATI peaks (energy peaks that do not vanish if $E_{780}$ is set to zero).  Now, $\alpha$ represents a distribution that is proportional to $\cos(\phi_{\mathrm{polar}}-\alpha-180\mathrm{\,deg})$ for sidebands and $\cos(\phi_{\mathrm{polar}}-\alpha)$ for ATI peaks.

\begin{figure}
\epsfig{file=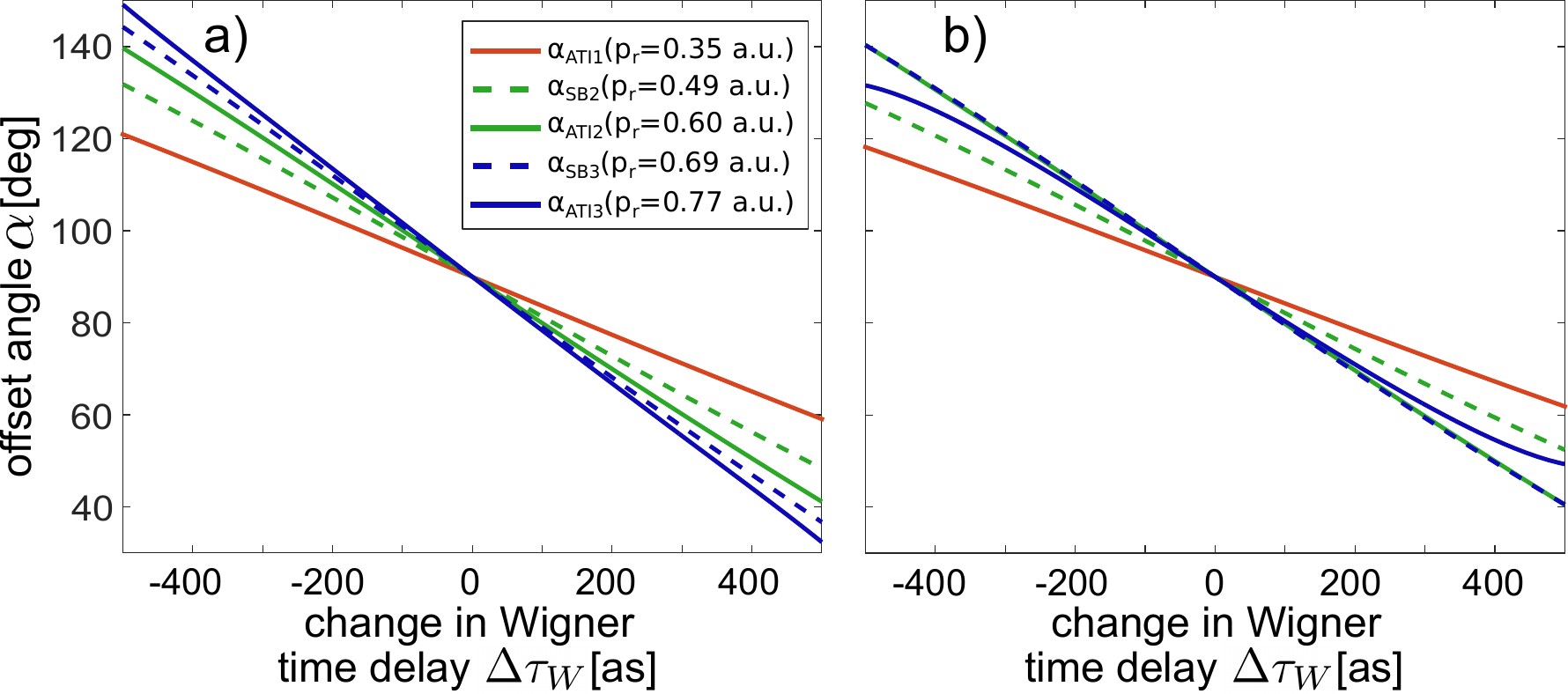, width=8.7cm}
\caption{(a) shows the change in Wigner time delay due to a linear phase of the initial momentum distribution for the case defined in Fig. \ref{fig_linearphasemomdistribution}(a). This relates the change in Wigner time delay (for a single-color field with $E_{390}=0.04$\,a.u.) and the offset angles $\alpha$ (that are obtained using a two-color field with $E_{390}=0.04$\,a.u. and $E_{780}=0.004$\,a.u.). (b) is analogous to (a) but uses the amplitude distribution from Fig. \ref{fig_linearphasemomdistribution}(b).}
\label{fig_figure9_} 
\end{figure}

Third, the offset angles $\alpha$ can be related to the derivative of the phase of the initial momentum distribution $\phi_\mathrm{off}^{\prime}$ using Fig. \ref{fig_linearphasemomdistribution}(e) as a look up table. From Eq. \ref{finalWT} the change in Wigner time delay, $\Delta \tau_W$, can be directly obtained for an offset angle $\alpha$. The result is presented in Fig. \ref{fig_figure9_}(a). However, this mapping of $\alpha$ to $\phi_\mathrm{off}^{\prime}$ and $\Delta \tau_W$ can only be done if the following conditions are fulfilled: (i) the laser parameters and the value for the ionization potential $I_p$ are as in this work, (ii) the envelope of the amplitudes can be neglected (see Fig. \ref{fig_linearphasemomdistribution}(f) and Fig. \ref{fig_figure9_}(b)), (iii) the sub-cycle dependence of the ionization rate \cite{Yudin2001A} can be neglected and (iv) Coulomb interaction after tunneling \cite{Bray2018} can be neglected. Conditions (i)-(iii) are not problematic because one could simply rerun the simulations by solving Eq. \ref{lab_initial} and Eq. \ref{lab1} using new laser parameters, the appropriate ionization potential and an amplitude distribution of the initial momentum distribution that does not only depend on the initial momentum ($B(p_i)$) but also on the electron release time ($B(p_i,t)$). This time-dependence could be obtained from theory \cite{Yudin2001A}. Alternatively, $B(p_i,t)$ could be estimated directly from the envelope of the measured electron momentum distribution (see e.g. Refs. \cite{Eckle2008,Ge2019,DanielArXiv2020}). In general, the measured electron momentum distribution should be normalized by the envelope of the measured electron momentum distribution in order to remove the modulation of the intensity from the measured interference pattern. An addiational benefit of the extracted envelope of the measured electron momentum distribution is that it allows one to estimate the absolute orientation of the laser electric field (even if Coulomb interaction after tunneling is not neglected \cite{Bray2018}).
Condition (iv) is usually not fulfilled. In section 8 it will be shown that for small values of $\alpha$, Coulomb interaction after tunneling just adds an additional, energy dependent offset angle, $\alpha_{\mathrm{Coulomb}}$, to the offset angle that is due to $\phi_\mathrm{off}^{\prime}$. Consequently, the measured offset angle, $\alpha$, can be expressed by $\alpha_{\mathrm{corrected}}=\alpha-\alpha_{\mathrm{Coulomb}}$ to be as precise as possible. An elegant alternative to circumvent problems regarding condition (iv) is to compare two different ionization channels (e.g. different kinetic energy releases for molecular dissociation) and assume that both have the same value $\alpha_{\mathrm{Coulomb}}$. The pairs of measured offset angles ($\alpha_{\mathrm{channel\,1}}$ and  $\alpha_{\mathrm{channel\,1}}$) can be used to calculate the difference in the Wigner time delay for the two channels. This is a good approximation if the mapping of $\Delta \tau_W$ to offset angle $\alpha$ is linear (which is the case for offset angles between $60\mathrm{\,deg}$ and $120\mathrm{\,deg}$). For this procedure the contribution of Coulomb interaction cancels out and allows for the experimental access of the difference in the Wigner time delay for the two ionization channels upon tunnel ionization.

\section{VIII. The influence of Coulomb Interaction after Tunneling}
In the previous sections, Coulomb interaction after tunneling of the electron with its parent ion was neglected. In the next step, we run a semi-classical two-step (SCTS) simulation (as in Refs. \cite{Shilovski2016,Eckart2018SubCycle}) which includes Coulomb interaction after tunneling. Comparison of the results obtained from the SCTS model with the results from the HASE model shows that Coulomb interaction after tunneling does not qualitatively change the obtained offset angles, $\alpha$, or the inferred values for the changes of the Wigner time delay, $\Delta\tau_W$. This indicates that - despite its simplicity - the HASE model captures the essential physics of the studied scenario. We emphasize that the SCTS model is not a full quantum simulation and that further experimental and theoretical benchmarks are needed. Refs. \cite{DanielArXiv2020,EckartArXivSideband} claim to provide experimental evidence that indicates that the HASE model is a good approximation.

\begin{figure}
\epsfig{file=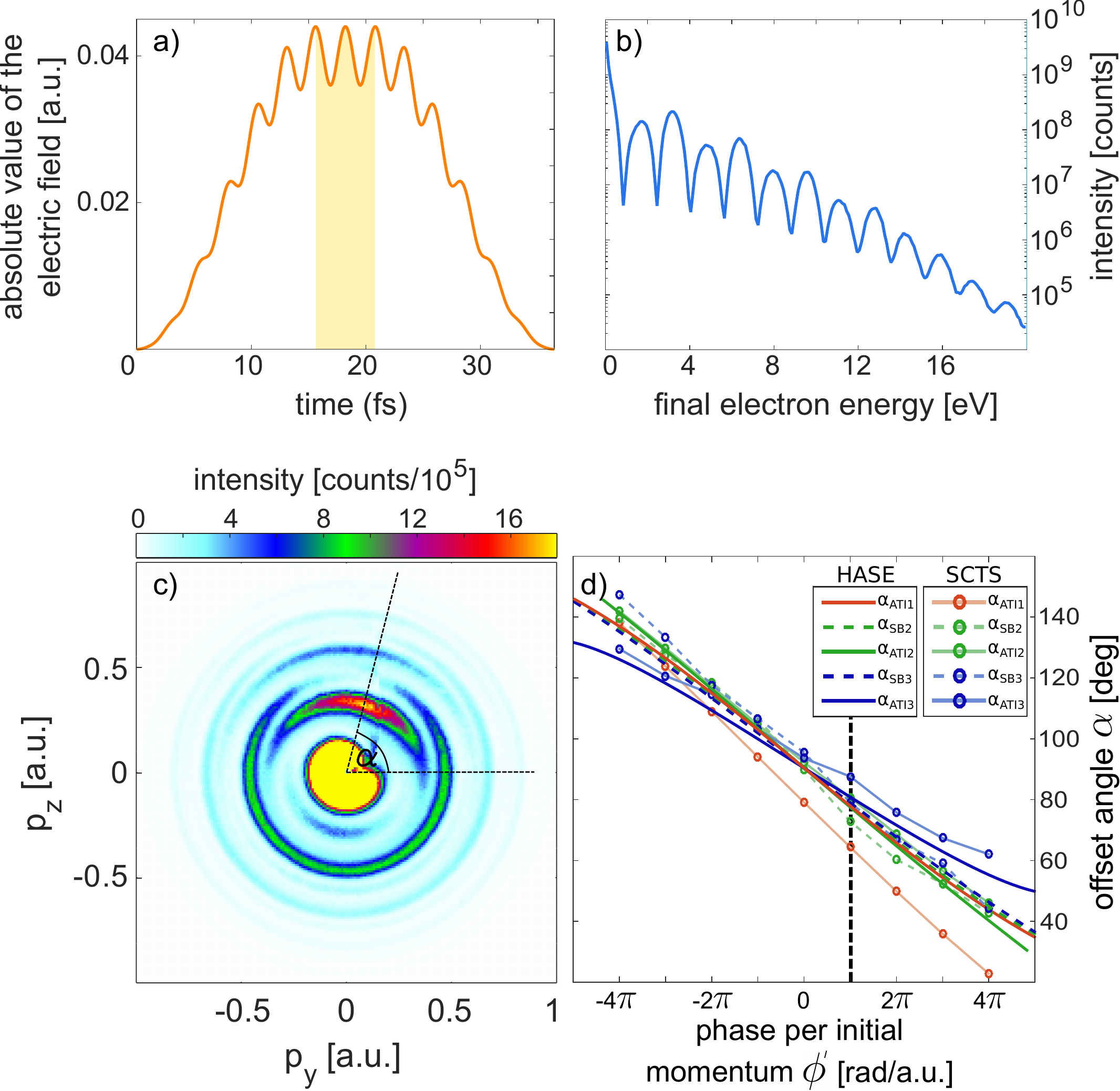, width=8.7cm}
\caption{(a) shows the absolute value of the time-dependent electric field used for the SCTS model. The shaded region indicates the electron release times that were considered for our SCTS model. (b) depicts the electron energy spectrum from the SCTS model. (c) presents the electron momentum distribution in the plane of polarization from the SCTS model with $\phi^{\prime}_{\mathrm{off}}=0$\,rad/a.u. (d) shows the offset angles, $\alpha$, as a function of $\phi^{\prime}_{\mathrm{off}}$ in full analogy to Fig. \ref{fig_linearphasemomdistribution}(f). The values of $\alpha$ that are obtained using the SCTS model are labeled with ``SCTS''. The data from Fig. \ref{fig_linearphasemomdistribution}(f) are shown for comparison and labeled with ``HASE''. $\phi^{\prime}_{\mathrm{off}}=\pi$\,rad/a.u. is indicated by a black dashed line.}
\label{fig_figure10label} 
\end{figure}

For the SCTS calculation, we use the same CoRTC field as in Fig. \ref{fig_figure1label}(a) and add an envelope with a total duration of 14 cycles of the light field ($14 T_{780}$). The rising and the falling edge of the laser pulse have a sine-square-shape and between, there is a flat envelope with a duration of 2 cycles of the light field ($2 T_{780}$). The absolute value of the electric field that is used for our SCTS calculation is shown in Fig. \ref{fig_figure10label}(a). To minimize the contributions of the rising and the falling edge of the laser pulse in our SCTS model, the electron release time is restricted to the inner two-cycles of the light pulse (shaded region in Fig. \ref{fig_figure10label}(a)). In contrast to the HASE model the SCTS model is a full three-dimensional model which does not neglect the initial momentum components in and against the light propagation direction. The momentum offset $\vec{p}_{\mathrm{0\_3D}}$ and the ionization probability that are used in the SCTS simulation (see \footnote{The tunneling probability for the SCTS model is assumed to be $R(\vec{p}_{\mathrm{i\_3D}})=\exp \left(-\frac{\left|\vec{p}_{\mathrm{i\_3D}}-\vec{p}_{\mathrm{0\_3D}}\right|^2}{2\sigma^2}\right)$, which is in full analogy to Fig. \ref{fig_linearphasemomdistribution}(b). The initial momentum at the tunnel exit is represented by $\vec{p}_{\mathrm{i\_3D}}=\begin{pmatrix}p_{\mathrm{0x}}\\p_{\mathrm{0\perp}}\\p_{\mathrm{0\parallel}}\end{pmatrix}$. Here, $\vec{p}_{\mathrm{i\_3D}}$ is defined using a reference frame that is aligned along the direction of the electric field, $\vec{E}(t_0)$, at the time the electron is released, $t_0$. Accordingly, $p_{\mathrm{0x}}$ points along the light propagation direction and $p_{\mathrm{0\parallel}}$ [$p_{\mathrm{0\perp}}$]  points along  the direction that is parallel (perpendicular) to $\vec{E}(t_0)$. The alignment of the direction that belongs to $p_{\mathrm{0\perp}}$ is chosen such that an increase in the value of $p_{\mathrm{0\perp}}$ leads to an increased absolute value of the final electron momentum (as for the HASE model, see discussion of Eq. \ref{lab_initial}). Here, we choose $\sigma=0.2$\,a.u., $p_{\mathrm{0\parallel}}=0$\,a.u. and $\vec{p}_{\mathrm{0\_3D}}=\begin{pmatrix}0\\0.2\\0\end{pmatrix}$ in full analogy to the HASE model. It should be noted, that due to the choice of $R(\vec{p}_{\mathrm{i\_3D}})$, the tunneling probability, does not depend on the absolute value of the electric field, which is a difference compared to the original SCTS model (see Eq. 9 from Ref. \cite{Shilovski2016}). }) are chosen to be very similar to those of the HASE model (see Fig. \ref{fig_linearphasemomdistribution}(b)). Using an ionization potential of $I_p=15.76$\,eV, we calculate 250 million semi-classical trajectories. The intensity in final electron momentum space is calculated as a coherent sum of all semi-classically modeled electrons using Eq. 15 from Ref. \cite{Shilovski2016} for a Cartesian three-dimensional grid in momentum space with a bin size of 0.01\,a.u. The resulting electron energy distribution is shown as a blue line in Fig. \ref{fig_figure10label}(b) and the projection of the electron momentum distribution to the light's polarization plane is shown in Fig. \ref{fig_figure10label}(c). Convergence of the SCTS simulation was achieved by implementing the method of ``phase compression'' as described in Ref. \cite{EckartDiss}. Convergence was verified by variation of the grid's bin size. It can be seen in Fig. \ref{fig_figure10label}(c) that the intensity of low energy electrons is very high (note that the color scale is saturated for radial momenta below 0.2\,a.u). This is not unexpected because the intensity of low-energy electrons  sensitively depends on the choice of $\vec{p}_{\mathrm{0\_3D}}$ \cite{Olga2011A}. For comparison with experimental data, $\vec{p}_{\mathrm{0\_3D}}$ could be chosen differently (e.g. using results from saddle-point strong field approximation \cite{Eckart2018_Offsets}) but the purpose of this section is to analyze the role of Coulomb interaction after tunneling by comparing the HASE model and the SCTS model. To this end the small radial electron momenta ($p_r<0.2$\,a.u.) are not important because we did not include them in the discussion of the HASE model at all.

\begin{figure}
\epsfig{file=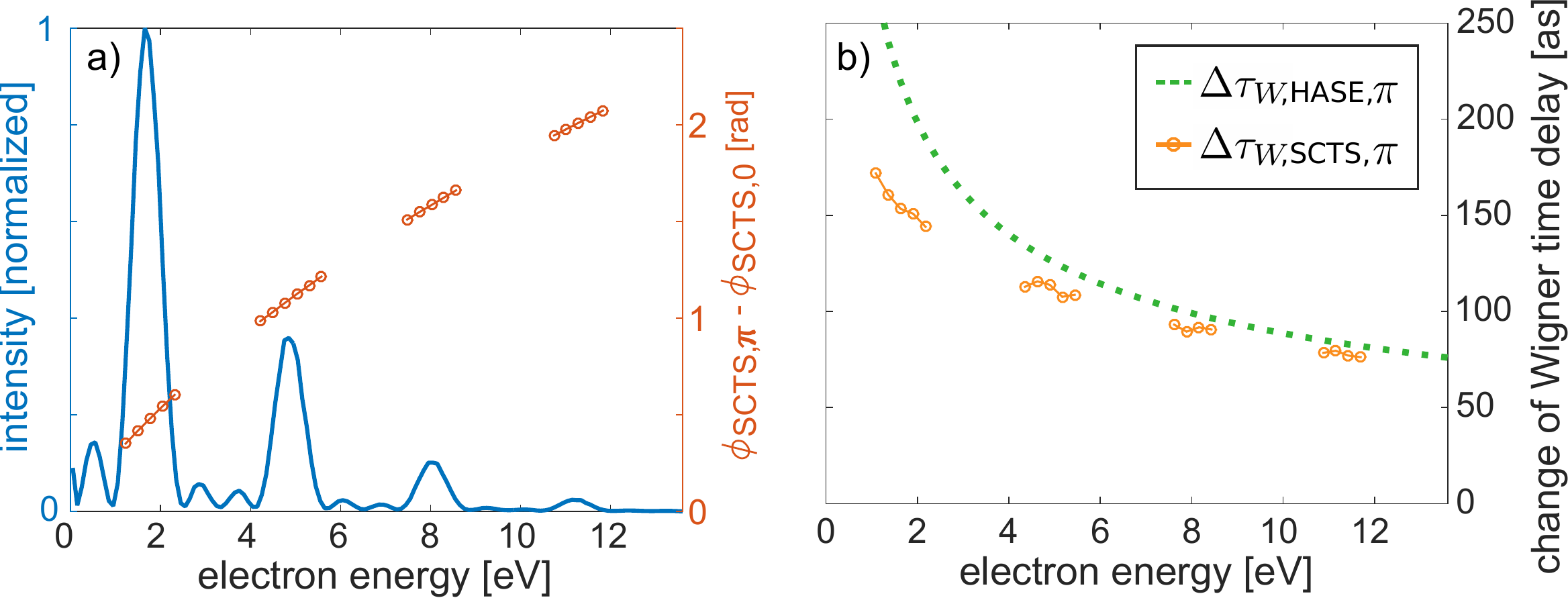, width=8.7cm}
\caption{The blue line in (a) shows the electron energy distribution that is obtained using the SCTS model for the single-color field ($E_{390}=0.04$\,a.u. and $E_{780}=0$\,a.u.). The change of the semi-classical phase, $\phi_{\mathrm{SCTS,}\pi}(E)-\phi_{\mathrm{SCTS,}0}(E)$, that results comparing an SCTS simulation with $\phi^{\prime}_{\mathrm{off}}=\pi$\,rad/a.u. and $\phi^{\prime}_{\mathrm{off}}=0$\,rad/a.u., is shown by the red circles. (b) shows the change of the Wigner time delay according to the SCTS model, $\Delta\tau_{W\mathrm{,SCTS,}\pi}$, as orange data points. It should be noted that $\Delta\tau_{W\mathrm{,SCTS,}\pi}$ is the derivative of the phase difference that is shown in (a). For comparison the Wigner time delay from the HASE model, $\Delta\tau_{W\mathrm{,HASE,}\pi}$, is shown as green dashed line.}
\label{fig_figure11label} 
\end{figure}

Fig. \ref{fig_figure10label}(c) is the first important result of the SCTS model with respect to the overall aim of this paper. Fig. \ref{fig_figure10label}(c) is obtained using the SCTS with no additional phase added to the trajectory at the tunnel exit (corresponds to $\phi^{\prime}_{\mathrm{off}}=0$\,rad/a.u.). In full analogy to the HASE model the SCTS model is extended adding the offset phase $\phi_{\mathrm{off}}(p_{\mathrm{0\perp}})=\kappa \frac{p_{\mathrm{0\perp}}}{a.u.}$. To compare the SCTS model and the HASE model, the value of $\phi^{\prime}_{\mathrm{off}}$ is systematically varied and the rotation angles, $\alpha$, are obtained as for Fig. \ref{fig_linearphasemomdistribution}(f). The results from the SCTS model and the results from the HASE model show excellent agreement (Fig. \ref{fig_figure10label}(d)). In particular, the peaks at medium electron energies ($\alpha_{SB2}$, $\alpha_{ATI2}$ and $\alpha_{SB3}$) only deviate by a few degrees for a wide range of $\phi^{\prime}_{\mathrm{off}}$ . For low and high electron energies ($\alpha_{ATI1}$ and $\alpha_{ATI3}$) deviations by up to $10\mathrm{\,deg}$ are found. Most importantly, it is evident that the slopes of all curves in Fig. \ref{fig_figure11label}(a) are similar. This is the first key-finding of this section because this constant slope is the necessary precondition that allows for the straight-forward mapping of changes of $\alpha$ to changes of the Wigner time delay (see sections 3-7).

The result in Fig. \ref{fig_figure10label}(d) shows that $\Delta \alpha$ can be used to determine $\Delta \phi^{\prime}_{\mathrm{off}}$. To further validate the HASE model, it remains to be shown that $\phi^{\prime}_{\mathrm{off}}$ allows one to infer $\Delta \tau_W$ as suggested in Eq. \ref{finalWT}. In the next step, the validity of Eq. \ref{finalWT} is tested using the SCTS model. To this end we have performed a second SCTS simulation using a single-color field ($E_{390}=0.04$\,a.u. and $E_{780}=0$\,a.u.) with all other parameters identical to the previous two-color simulation and excluded recolliding trajectories (characterized by reaching a minimal distance below 10\,a.u. to the nucleus, this is the case for about 5\% of all calculated trajectories). The resulting electron energy spectrum is shown as a blue line in Fig. \ref{fig_figure11label}(a). 

The SCTS calculation does not only give access to the absolute square of the semi-classically modeled wave function (see Eq. 15 from Ref. \cite{Shilovski2016}) but can also be used to investigate the phase of the semi-classically modeled wave function (using the argument of the expression between the dashes in Eq. 15 from Ref. \cite{Shilovski2016}). For a thin slice along the light propagation direction $-0.01\mathrm{\,a.u.}<p_x<0.01\mathrm{\,a.u.}$ the change of the semi-classical phase, $\phi_{\mathrm{SCTS,}\pi}-\phi_{\mathrm{SCTS,}0}$, is investigated. Here, $\phi_{\mathrm{SCTS,}\pi}$ [$\phi_{\mathrm{SCTS,}0}$] is the phase from the SCTS model using $\phi^{\prime}_{\mathrm{off}}=\pi$\,rad/a.u. [$\phi^{\prime}_{\mathrm{off}}=0$\,rad/a.u.]. It is found that  $\phi_{\mathrm{SCTS,}\pi}-\phi_{\mathrm{SCTS,}0}$ only depends on the electron's energy, $E$, which is expected due to the symmetry of circularly polarized light. The energy dependent value for $\phi_{\mathrm{SCTS,}\pi}-\phi_{\mathrm{SCTS,}0}$ is shown in Fig. \ref{fig_figure11label}(a) in red. This phase difference allows one to express the corresponding change of the Wigner time delay using the SCTS model by:
\begin{equation}
\begin{aligned}
\Delta&\tau_{W\mathrm{,SCTS,}\pi}(E) \\
&=\hbar \frac{\partial \phi_{\mathrm{SCTS,}\pi}(E)}{\partial E}-\hbar\frac{\partial\phi_{\mathrm{SCTS,}0}(E)}{\partial E}\\
&=\hbar \frac{\partial (\phi_{\mathrm{SCTS,}\pi}(E)-\phi_{\mathrm{SCTS,}0}(E))}{\partial E}
\label{WignerSCTS}
\end{aligned}
\end{equation}

Eq. \ref{WignerSCTS} is evaluated to obtain $\Delta\tau_{W\mathrm{,SCTS,}\pi}(E)$, which is shown in Fig. \ref{fig_figure11label}(b). In order to compare $\Delta\tau_{W\mathrm{,SCTS,}\pi}(E)$ to the results from the HASE model, the change of the Wigner time delay according to the HASE model is expressed by (see Eq. \ref{finalWT}):
\begin{equation}
\Delta \tau_{\mathrm{W,HASE,}\pi}(E)=\hbar \frac{m_e}{\sqrt{2 E}}\frac{\pi\mathrm{\,rad}}{\mathrm{a.u.}}
\end{equation}
The result for $\Delta\tau_{W\mathrm{,HASE,}\pi}(E)$ is shown as green dashed line in Fig. \ref{fig_figure11label}(b). The obtained values for $\Delta\tau_W$ from the SCTS and the HASE model qualitatively agree. For high electron energies the agreement is very good and for low electron energies Coulomb interaction after tunneling reduces $\Delta\tau_W$ by about 30\% in the SCTS model compared to the HASE model. This allows one to conclude that the mapping of $\phi^{\prime}_{\mathrm{off}}=\pi$\,rad/a.u. to changes of the Wigner time delay qualitatively agrees for the HASE model and the SCTS model. As expected, the deviations for the first ATI peak are larger than those of the other ATI peaks because Coulomb interaction after tunneling has a stronger impact on slow electrons than it has on fast electrons. 
 
Finally, it should be noted that the choice of $\phi^{\prime}_{\mathrm{off}}=\pi$\,rad/a.u. is chosen without loss of generality because the Wigner time delays shown in Fig. \ref{fig_figure11label}(b) are expected to scale linearly with $\phi^{\prime}_{\mathrm{off}}$. The fact, that the mapping of $\phi^{\prime}_{\mathrm{off}}$ to $\Delta \tau_W$ (see Eq. \ref{finalWT}) is qualitatively reproduced by the SCTS model is the second key-finding of this section.

\section{IX. Conclusion}
Building on the simplified HASE model, it is shown that the angular distribution of main ATI peaks and sidebands can be used to infer the derivative of the phase of the initial momentum distribution, $\phi_\mathrm{off}^{\prime}$, from experimentally accessible quantities. Further, it is found that $\phi_\mathrm{off}^{\prime}$ can be related to changes of the Wigner time delay $\Delta \tau_W$ within the HASE model. Finally, an intuitive interpretation of the Wigner time delay for tunnel ionization is suggested that links the Wigner time delay to a displacement of the wave function in position space. The findings from the HASE model are compared to results from an SCTS model that includes Coulomb interaction after tunneling.

In future (coincidence) experiments, the offset angles $\alpha$ of ATI peaks and sidebands could be examined as a function of the kinetic energy release for molecular dissociation, the molecular orientation or the atomic species. Recently, it has been claimed that the angular dependence of the Wigner time delay upon tunnel ionization of molecular hydrogen has been measured using HASE \cite{DanielArXiv2020}. In another recent work, it has been suggested that the HASE model is not limited to ionization in co-rotating two-color fields but can also be used to model ionization in counter-rotating two-color fields \cite{EckartArXivSideband}. In conclusion, HASE paves the road towards the measurement of molecular structure, polarization states and non-adiabatic fingerprints of tunnel ionization in position space with sub-cycle temporal resolution.

\begin{acknowledgments}
\section{acknowledgments}
\normalsize
This work was funded by the German Research Foundation (DFG) through the Priority Programme Quantum Dynamics in Tailored Intense Fields (SPP 1840). Thanks to Daniel Trabert, Kilian Fehre, Alexander Hartung, Jonas Rist, Angelina Geyer, Maksim Kunitski, Horst Schmidt-B\"ocking, Lothar Ph. H. Schmidt, Till Jahnke and Reinhard D\"orner for fruitful discussions and comments regarding the manuscript. The SCTS calculations were done using a modified version of a code that was originally developed by Maksim Kunitski and is based on Ref. \cite{Shilovski2016}.
\end{acknowledgments}

\bibliographystyle{apsrev4-1}

\begin{thebibliography}{44}%
\makeatletter
\providecommand \@ifxundefined [1]{%
 \@ifx{#1\undefined}
}%
\providecommand \@ifnum [1]{%
 \ifnum #1\expandafter \@firstoftwo
 \else \expandafter \@secondoftwo
 \fi
}%
\providecommand \@ifx [1]{%
 \ifx #1\expandafter \@firstoftwo
 \else \expandafter \@secondoftwo
 \fi
}%
\providecommand \natexlab [1]{#1}%
\providecommand \enquote  [1]{``#1''}%
\providecommand \bibnamefont  [1]{#1}%
\providecommand \bibfnamefont [1]{#1}%
\providecommand \citenamefont [1]{#1}%
\providecommand \href@noop [0]{\@secondoftwo}%
\providecommand \href [0]{\begingroup \@sanitize@url \@href}%
\providecommand \@href[1]{\@@startlink{#1}\@@href}%
\providecommand \@@href[1]{\endgroup#1\@@endlink}%
\providecommand \@sanitize@url [0]{\catcode `\\12\catcode `\$12\catcode
  `\&12\catcode `\#12\catcode `\^12\catcode `\_12\catcode `\%12\relax}%
\providecommand \@@startlink[1]{}%
\providecommand \@@endlink[0]{}%
\providecommand \url  [0]{\begingroup\@sanitize@url \@url }%
\providecommand \@url [1]{\endgroup\@href {#1}{\urlprefix }}%
\providecommand \urlprefix  [0]{URL }%
\providecommand \Eprint [0]{\href }%
\providecommand \doibase [0]{http://dx.doi.org/}%
\providecommand \selectlanguage [0]{\@gobble}%
\providecommand \bibinfo  [0]{\@secondoftwo}%
\providecommand \bibfield  [0]{\@secondoftwo}%
\providecommand \translation [1]{[#1]}%
\providecommand \BibitemOpen [0]{}%
\providecommand \bibitemStop [0]{}%
\providecommand \bibitemNoStop [0]{.\EOS\space}%
\providecommand \EOS [0]{\spacefactor3000\relax}%
\providecommand \BibitemShut  [1]{\csname bibitem#1\endcsname}%
\let\auto@bib@innerbib\@empty
\bibitem [{\citenamefont {Keldysh}(1965)}]{Keldysh1965}%
  \BibitemOpen
  \bibfield  {author} {\bibinfo {author} {\bibfnamefont {L.~V.}\ \bibnamefont
  {Keldysh}},\ }\href@noop {} {\bibfield  {journal} {\bibinfo  {journal} {Sov.
  Phys. JETP}\ }\textbf {\bibinfo {volume} {20}},\ \bibinfo {pages} {1307}
  (\bibinfo {year} {1965})}\BibitemShut {NoStop}%
\bibitem [{\citenamefont {Voronov}\ and\ \citenamefont
  {Delone}(1966)}]{voronov1966many}%
  \BibitemOpen
  \bibfield  {author} {\bibinfo {author} {\bibfnamefont {G.~S.}\ \bibnamefont
  {Voronov}}\ and\ \bibinfo {author} {\bibfnamefont {N.~B.}\ \bibnamefont
  {Delone}},\ }\href@noop {} {\bibfield  {journal} {\bibinfo  {journal} {Sov.
  Phys. JETP}\ }\textbf {\bibinfo {volume} {23}},\ \bibinfo {pages} {54}
  (\bibinfo {year} {1966})}\BibitemShut {NoStop}%
\bibitem [{\citenamefont {Freeman}\ \emph {et~al.}(1987)\citenamefont
  {Freeman}, \citenamefont {Bucksbaum}, \citenamefont {Milchberg},
  \citenamefont {Darack}, \citenamefont {Schumacher},\ and\ \citenamefont
  {Geusic}}]{Freeman1987}%
  \BibitemOpen
  \bibfield  {author} {\bibinfo {author} {\bibfnamefont {R.~R.}\ \bibnamefont
  {Freeman}}, \bibinfo {author} {\bibfnamefont {P.~H.}\ \bibnamefont
  {Bucksbaum}}, \bibinfo {author} {\bibfnamefont {H.}~\bibnamefont
  {Milchberg}}, \bibinfo {author} {\bibfnamefont {S.}~\bibnamefont {Darack}},
  \bibinfo {author} {\bibfnamefont {D.}~\bibnamefont {Schumacher}}, \ and\
  \bibinfo {author} {\bibfnamefont {M.~E.}\ \bibnamefont {Geusic}},\ }\href
  {\doibase 10.1103/PhysRevLett.59.1092} {\bibfield  {journal} {\bibinfo
  {journal} {Phys. Rev. Lett.}\ }\textbf {\bibinfo {volume} {59}},\ \bibinfo
  {pages} {1092} (\bibinfo {year} {1987})}\BibitemShut {NoStop}%
\bibitem [{\citenamefont {{{M. Yu. Ivanov}}}\ \emph {et~al.}(2005)\citenamefont
  {{{M. Yu. Ivanov}}}, \citenamefont {Spanner},\ and\ \citenamefont
  {Smirnova}}]{Misha2005}%
  \BibitemOpen
  \bibfield  {author} {\bibinfo {author} {\bibnamefont {{{M. Yu. Ivanov}}}},
  \bibinfo {author} {\bibfnamefont {M.}~\bibnamefont {Spanner}}, \ and\
  \bibinfo {author} {\bibfnamefont {O.}~\bibnamefont {Smirnova}},\ }\href
  {\doibase 10.1080/0950034042000275360} {\bibfield  {journal} {\bibinfo
  {journal} {J. Mod. Opt.}\ }\textbf {\bibinfo {volume} {52}},\ \bibinfo
  {pages} {165} (\bibinfo {year} {2005})}\BibitemShut {NoStop}%
\bibitem [{\citenamefont {Einstein}(1905)}]{Einstein1905}%
  \BibitemOpen
  \bibfield  {author} {\bibinfo {author} {\bibfnamefont {A.}~\bibnamefont
  {Einstein}},\ }\href {\doibase 10.1002/andp.19053220607} {\bibfield
  {journal} {\bibinfo  {journal} {Ann. Phys.}\ }\textbf {\bibinfo {volume}
  {322}},\ \bibinfo {pages} {132} (\bibinfo {year} {1905})}\BibitemShut
  {NoStop}%
\bibitem [{\citenamefont {Arb\'o}\ \emph {et~al.}(2010)\citenamefont {Arb\'o},
  \citenamefont {Ishikawa}, \citenamefont {Schiessl}, \citenamefont {Persson},\
  and\ \citenamefont {Burgd\"orfer}}]{Arbo2010}%
  \BibitemOpen
  \bibfield  {author} {\bibinfo {author} {\bibfnamefont {D.~G.}\ \bibnamefont
  {Arb\'o}}, \bibinfo {author} {\bibfnamefont {K.~L.}\ \bibnamefont
  {Ishikawa}}, \bibinfo {author} {\bibfnamefont {K.}~\bibnamefont {Schiessl}},
  \bibinfo {author} {\bibfnamefont {E.}~\bibnamefont {Persson}}, \ and\
  \bibinfo {author} {\bibfnamefont {J.}~\bibnamefont {Burgd\"orfer}},\ }\href
  {\doibase 10.1103/PhysRevA.81.021403} {\bibfield  {journal} {\bibinfo
  {journal} {Phys. Rev. A}\ }\textbf {\bibinfo {volume} {81}},\ \bibinfo
  {pages} {021403(R)} (\bibinfo {year} {2010})}\BibitemShut {NoStop}%
\bibitem [{\citenamefont {Zipp}\ \emph {et~al.}(2014)\citenamefont {Zipp},
  \citenamefont {Natan},\ and\ \citenamefont {Bucksbaum}}]{Zipp2014}%
  \BibitemOpen
  \bibfield  {author} {\bibinfo {author} {\bibfnamefont {L.~J.}\ \bibnamefont
  {Zipp}}, \bibinfo {author} {\bibfnamefont {A.}~\bibnamefont {Natan}}, \ and\
  \bibinfo {author} {\bibfnamefont {P.~H.}\ \bibnamefont {Bucksbaum}},\ }\href
  {\doibase 10.1364/OPTICA.1.000361} {\bibfield  {journal} {\bibinfo  {journal}
  {Optica}\ }\textbf {\bibinfo {volume} {1}},\ \bibinfo {pages} {361} (\bibinfo
  {year} {2014})}\BibitemShut {NoStop}%
\bibitem [{\citenamefont {Han}\ \emph {et~al.}(2018)\citenamefont {Han},
  \citenamefont {Ge}, \citenamefont {Shao}, \citenamefont {Gong},\ and\
  \citenamefont {Liu}}]{Han2018}%
  \BibitemOpen
  \bibfield  {author} {\bibinfo {author} {\bibfnamefont {M.}~\bibnamefont
  {Han}}, \bibinfo {author} {\bibfnamefont {P.}~\bibnamefont {Ge}}, \bibinfo
  {author} {\bibfnamefont {Y.}~\bibnamefont {Shao}}, \bibinfo {author}
  {\bibfnamefont {Q.}~\bibnamefont {Gong}}, \ and\ \bibinfo {author}
  {\bibfnamefont {Y.}~\bibnamefont {Liu}},\ }\href {\doibase
  10.1103/PhysRevLett.120.073202} {\bibfield  {journal} {\bibinfo  {journal}
  {Phys. Rev. Lett.}\ }\textbf {\bibinfo {volume} {120}},\ \bibinfo {pages}
  {073202} (\bibinfo {year} {2018})}\BibitemShut {NoStop}%
\bibitem [{\citenamefont {Ge}\ \emph {et~al.}(2019)\citenamefont {Ge},
  \citenamefont {Han}, \citenamefont {Deng}, \citenamefont {Gong},\ and\
  \citenamefont {Liu}}]{Ge2019}%
  \BibitemOpen
  \bibfield  {author} {\bibinfo {author} {\bibfnamefont {P.}~\bibnamefont
  {Ge}}, \bibinfo {author} {\bibfnamefont {M.}~\bibnamefont {Han}}, \bibinfo
  {author} {\bibfnamefont {Y.}~\bibnamefont {Deng}}, \bibinfo {author}
  {\bibfnamefont {Q.}~\bibnamefont {Gong}}, \ and\ \bibinfo {author}
  {\bibfnamefont {Y.}~\bibnamefont {Liu}},\ }\href {\doibase
  10.1103/PhysRevLett.122.013201} {\bibfield  {journal} {\bibinfo  {journal}
  {Phys. Rev. Lett.}\ }\textbf {\bibinfo {volume} {122}},\ \bibinfo {pages}
  {013201} (\bibinfo {year} {2019})}\BibitemShut {NoStop}%
\bibitem [{\citenamefont {Feng}\ \emph {et~al.}(2019)\citenamefont {Feng},
  \citenamefont {Li}, \citenamefont {Luo}, \citenamefont {Liu}, \citenamefont
  {Du}, \citenamefont {Zhou},\ and\ \citenamefont {Lu}}]{Feng2019}%
  \BibitemOpen
  \bibfield  {author} {\bibinfo {author} {\bibfnamefont {Y.}~\bibnamefont
  {Feng}}, \bibinfo {author} {\bibfnamefont {M.}~\bibnamefont {Li}}, \bibinfo
  {author} {\bibfnamefont {S.}~\bibnamefont {Luo}}, \bibinfo {author}
  {\bibfnamefont {K.}~\bibnamefont {Liu}}, \bibinfo {author} {\bibfnamefont
  {B.}~\bibnamefont {Du}}, \bibinfo {author} {\bibfnamefont {Y.}~\bibnamefont
  {Zhou}}, \ and\ \bibinfo {author} {\bibfnamefont {P.}~\bibnamefont {Lu}},\
  }\href {\doibase 10.1103/PhysRevA.100.063411} {\bibfield  {journal} {\bibinfo
   {journal} {Phys. Rev. A}\ }\textbf {\bibinfo {volume} {100}},\ \bibinfo
  {pages} {063411} (\bibinfo {year} {2019})}\BibitemShut {NoStop}%
\bibitem [{\citenamefont {Huismans}\ \emph {et~al.}(2011)\citenamefont
  {Huismans}, \citenamefont {Rouz{\'e}e}, \citenamefont {Gijsbertsen},
  \citenamefont {Jungmann}, \citenamefont {Smolkowska}, \citenamefont {Logman},
  \citenamefont {L{\'e}pine}, \citenamefont {Cauchy}, \citenamefont {Zamith},
  \citenamefont {Marchenko}, \citenamefont {Bakker}, \citenamefont {Berden},
  \citenamefont {Redlich}, \citenamefont {van~der Meer}, \citenamefont
  {Muller}, \citenamefont {Vermin}, \citenamefont {Schafer}, \citenamefont
  {Spanner}, \citenamefont {Ivanov}, \citenamefont {Smirnova}, \citenamefont
  {Bauer}, \citenamefont {Popruzhenko},\ and\ \citenamefont
  {Vrakking}}]{Huismans2011}%
  \BibitemOpen
  \bibfield  {author} {\bibinfo {author} {\bibfnamefont {Y.}~\bibnamefont
  {Huismans}}, \bibinfo {author} {\bibfnamefont {A.}~\bibnamefont
  {Rouz{\'e}e}}, \bibinfo {author} {\bibfnamefont {A.}~\bibnamefont
  {Gijsbertsen}}, \bibinfo {author} {\bibfnamefont {J.~H.}\ \bibnamefont
  {Jungmann}}, \bibinfo {author} {\bibfnamefont {A.~S.}\ \bibnamefont
  {Smolkowska}}, \bibinfo {author} {\bibfnamefont {P.~S. W.~M.}\ \bibnamefont
  {Logman}}, \bibinfo {author} {\bibfnamefont {F.}~\bibnamefont {L{\'e}pine}},
  \bibinfo {author} {\bibfnamefont {C.}~\bibnamefont {Cauchy}}, \bibinfo
  {author} {\bibfnamefont {S.}~\bibnamefont {Zamith}}, \bibinfo {author}
  {\bibfnamefont {T.}~\bibnamefont {Marchenko}}, \bibinfo {author}
  {\bibfnamefont {J.~M.}\ \bibnamefont {Bakker}}, \bibinfo {author}
  {\bibfnamefont {G.}~\bibnamefont {Berden}}, \bibinfo {author} {\bibfnamefont
  {B.}~\bibnamefont {Redlich}}, \bibinfo {author} {\bibfnamefont {A.~F.~G.}\
  \bibnamefont {van~der Meer}}, \bibinfo {author} {\bibfnamefont {H.~G.}\
  \bibnamefont {Muller}}, \bibinfo {author} {\bibfnamefont {W.}~\bibnamefont
  {Vermin}}, \bibinfo {author} {\bibfnamefont {K.~J.}\ \bibnamefont {Schafer}},
  \bibinfo {author} {\bibfnamefont {M.}~\bibnamefont {Spanner}}, \bibinfo
  {author} {\bibfnamefont {M.~Y.}\ \bibnamefont {Ivanov}}, \bibinfo {author}
  {\bibfnamefont {O.}~\bibnamefont {Smirnova}}, \bibinfo {author}
  {\bibfnamefont {D.}~\bibnamefont {Bauer}}, \bibinfo {author} {\bibfnamefont
  {S.~V.}\ \bibnamefont {Popruzhenko}}, \ and\ \bibinfo {author} {\bibfnamefont
  {M.~J.~J.}\ \bibnamefont {Vrakking}},\ }\href {\doibase
  10.1126/science.1198450} {\bibfield  {journal} {\bibinfo  {journal}
  {Science}\ }\textbf {\bibinfo {volume} {331}},\ \bibinfo {pages} {61}
  (\bibinfo {year} {2011})}\BibitemShut {NoStop}%
\bibitem [{\citenamefont {Eckart}\ \emph
  {et~al.}(2018{\natexlab{a}})\citenamefont {Eckart}, \citenamefont {Kunitski},
  \citenamefont {Ivanov}, \citenamefont {Richter}, \citenamefont {Fehre},
  \citenamefont {Hartung}, \citenamefont {Rist}, \citenamefont {Henrichs},
  \citenamefont {Trabert}, \citenamefont {Schlott}, \citenamefont {{{L. Ph. H.
  Schmidt}}}, \citenamefont {Jahnke}, \citenamefont {Sch\"offler},
  \citenamefont {Kheifets},\ and\ \citenamefont
  {D\"orner}}]{Eckart2018SubCycle}%
  \BibitemOpen
  \bibfield  {author} {\bibinfo {author} {\bibfnamefont {S.}~\bibnamefont
  {Eckart}}, \bibinfo {author} {\bibfnamefont {M.}~\bibnamefont {Kunitski}},
  \bibinfo {author} {\bibfnamefont {I.}~\bibnamefont {Ivanov}}, \bibinfo
  {author} {\bibfnamefont {M.}~\bibnamefont {Richter}}, \bibinfo {author}
  {\bibfnamefont {K.}~\bibnamefont {Fehre}}, \bibinfo {author} {\bibfnamefont
  {A.}~\bibnamefont {Hartung}}, \bibinfo {author} {\bibfnamefont
  {J.}~\bibnamefont {Rist}}, \bibinfo {author} {\bibfnamefont {K.}~\bibnamefont
  {Henrichs}}, \bibinfo {author} {\bibfnamefont {D.}~\bibnamefont {Trabert}},
  \bibinfo {author} {\bibfnamefont {N.}~\bibnamefont {Schlott}}, \bibinfo
  {author} {\bibnamefont {{{L. Ph. H. Schmidt}}}}, \bibinfo {author}
  {\bibfnamefont {T.}~\bibnamefont {Jahnke}}, \bibinfo {author} {\bibfnamefont
  {M.~S.}\ \bibnamefont {Sch\"offler}}, \bibinfo {author} {\bibfnamefont
  {A.}~\bibnamefont {Kheifets}}, \ and\ \bibinfo {author} {\bibfnamefont
  {R.}~\bibnamefont {D\"orner}},\ }\href {\doibase 10.1103/PhysRevA.97.041402}
  {\bibfield  {journal} {\bibinfo  {journal} {Phys. Rev. A}\ }\textbf {\bibinfo
  {volume} {97}},\ \bibinfo {pages} {041402(R)} (\bibinfo {year}
  {2018}{\natexlab{a}})}\BibitemShut {NoStop}%
\bibitem [{\citenamefont {He}\ \emph {et~al.}(2018)\citenamefont {He},
  \citenamefont {Li}, \citenamefont {Zhou}, \citenamefont {Li}, \citenamefont
  {Cao},\ and\ \citenamefont {Lu}}]{ValenceElectronMotion2018}%
  \BibitemOpen
  \bibfield  {author} {\bibinfo {author} {\bibfnamefont {M.}~\bibnamefont
  {He}}, \bibinfo {author} {\bibfnamefont {Y.}~\bibnamefont {Li}}, \bibinfo
  {author} {\bibfnamefont {Y.}~\bibnamefont {Zhou}}, \bibinfo {author}
  {\bibfnamefont {M.}~\bibnamefont {Li}}, \bibinfo {author} {\bibfnamefont
  {W.}~\bibnamefont {Cao}}, \ and\ \bibinfo {author} {\bibfnamefont
  {P.}~\bibnamefont {Lu}},\ }\href {\doibase 10.1103/PhysRevLett.120.133204}
  {\bibfield  {journal} {\bibinfo  {journal} {Phys. Rev. Lett.}\ }\textbf
  {\bibinfo {volume} {120}},\ \bibinfo {pages} {133204} (\bibinfo {year}
  {2018})}\BibitemShut {NoStop}%
\bibitem [{\citenamefont {Wigner}(1955)}]{Wigner1955}%
  \BibitemOpen
  \bibfield  {author} {\bibinfo {author} {\bibfnamefont {E.~P.}\ \bibnamefont
  {Wigner}},\ }\href {\doibase 10.1103/PhysRev.98.145} {\bibfield  {journal}
  {\bibinfo  {journal} {Phys. Rev.}\ }\textbf {\bibinfo {volume} {98}},\
  \bibinfo {pages} {145} (\bibinfo {year} {1955})}\BibitemShut {NoStop}%
\bibitem [{\citenamefont {Pazourek}\ \emph {et~al.}(2015)\citenamefont
  {Pazourek}, \citenamefont {Nagele},\ and\ \citenamefont
  {Burgd\"orfer}}]{Pazourek2015}%
  \BibitemOpen
  \bibfield  {author} {\bibinfo {author} {\bibfnamefont {R.}~\bibnamefont
  {Pazourek}}, \bibinfo {author} {\bibfnamefont {S.}~\bibnamefont {Nagele}}, \
  and\ \bibinfo {author} {\bibfnamefont {J.}~\bibnamefont {Burgd\"orfer}},\
  }\href {\doibase 10.1103/RevModPhys.87.765} {\bibfield  {journal} {\bibinfo
  {journal} {Rev. Mod. Phys.}\ }\textbf {\bibinfo {volume} {87}},\ \bibinfo
  {pages} {765} (\bibinfo {year} {2015})}\BibitemShut {NoStop}%
\bibitem [{\citenamefont {Eppink}\ and\ \citenamefont
  {Parker}(1997)}]{Eppink1997}%
  \BibitemOpen
  \bibfield  {author} {\bibinfo {author} {\bibfnamefont {A.~T. J.~B.}\
  \bibnamefont {Eppink}}\ and\ \bibinfo {author} {\bibfnamefont {D.~H.}\
  \bibnamefont {Parker}},\ }\href {\doibase 10.1063/1.1148310} {\bibfield
  {journal} {\bibinfo  {journal} {Review of Scientific Instruments}\ }\textbf
  {\bibinfo {volume} {68}},\ \bibinfo {pages} {3477} (\bibinfo {year}
  {1997})}\BibitemShut {NoStop}%
\bibitem [{\citenamefont {Jagutzki}\ \emph {et~al.}(2002)\citenamefont
  {Jagutzki}, \citenamefont {Cerezo}, \citenamefont {Czasch}, \citenamefont
  {D{\"o}rner}, \citenamefont {Hattas}, \citenamefont {Huang}, \citenamefont
  {Mergel}, \citenamefont {Spillmann}, \citenamefont {Ullmann-Pfleger},
  \citenamefont {Weber}, \citenamefont {Schmidt-B{\"o}cking},\ and\
  \citenamefont {Smith}}]{jagutzki2002multiple}%
  \BibitemOpen
  \bibfield  {author} {\bibinfo {author} {\bibfnamefont {O.}~\bibnamefont
  {Jagutzki}}, \bibinfo {author} {\bibfnamefont {A.}~\bibnamefont {Cerezo}},
  \bibinfo {author} {\bibfnamefont {A.}~\bibnamefont {Czasch}}, \bibinfo
  {author} {\bibfnamefont {R.}~\bibnamefont {D{\"o}rner}}, \bibinfo {author}
  {\bibfnamefont {M.}~\bibnamefont {Hattas}}, \bibinfo {author} {\bibfnamefont
  {M.}~\bibnamefont {Huang}}, \bibinfo {author} {\bibfnamefont
  {V.}~\bibnamefont {Mergel}}, \bibinfo {author} {\bibfnamefont
  {U.}~\bibnamefont {Spillmann}}, \bibinfo {author} {\bibfnamefont
  {K.}~\bibnamefont {Ullmann-Pfleger}}, \bibinfo {author} {\bibfnamefont
  {T.}~\bibnamefont {Weber}}, \bibinfo {author} {\bibfnamefont
  {H.}~\bibnamefont {Schmidt-B{\"o}cking}}, \ and\ \bibinfo {author}
  {\bibfnamefont {G.~D.~W.}\ \bibnamefont {Smith}},\ }\href@noop {} {\bibfield
  {journal} {\bibinfo  {journal} {IEEE Trans. Nucl. Sci.}\ }\textbf {\bibinfo
  {volume} {49}},\ \bibinfo {pages} {2477} (\bibinfo {year}
  {2002})}\BibitemShut {NoStop}%
\bibitem [{\citenamefont {Ullrich}\ \emph {et~al.}(2003)\citenamefont
  {Ullrich}, \citenamefont {Moshammer}, \citenamefont {Dorn}, \citenamefont
  {D{\"o}rner}, \citenamefont {{{L. Ph. H. Schmidt}}},\ and\ \citenamefont
  {Schmidt-B{\"o}cking}}]{ullrich2003recoil}%
  \BibitemOpen
  \bibfield  {author} {\bibinfo {author} {\bibfnamefont {J.}~\bibnamefont
  {Ullrich}}, \bibinfo {author} {\bibfnamefont {R.}~\bibnamefont {Moshammer}},
  \bibinfo {author} {\bibfnamefont {A.}~\bibnamefont {Dorn}}, \bibinfo {author}
  {\bibfnamefont {R.}~\bibnamefont {D{\"o}rner}}, \bibinfo {author}
  {\bibnamefont {{{L. Ph. H. Schmidt}}}}, \ and\ \bibinfo {author}
  {\bibfnamefont {H.}~\bibnamefont {Schmidt-B{\"o}cking}},\ }\href@noop {}
  {\bibfield  {journal} {\bibinfo  {journal} {Rep. Prog. Phys.}\ }\textbf
  {\bibinfo {volume} {66}},\ \bibinfo {pages} {1463} (\bibinfo {year}
  {2003})}\BibitemShut {NoStop}%
\bibitem [{\citenamefont {Shvetsov-Shilovski}\ \emph
  {et~al.}(2016)\citenamefont {Shvetsov-Shilovski}, \citenamefont {Lein},
  \citenamefont {Madsen}, \citenamefont {R\"as\"anen}, \citenamefont {Lemell},
  \citenamefont {Burgd\"orfer}, \citenamefont {Arb\'o},\ and\ \citenamefont
  {T\ifmmode\mbox{\H{o}}\else\H{o}\fi{}k\'esi}}]{Shilovski2016}%
  \BibitemOpen
  \bibfield  {author} {\bibinfo {author} {\bibfnamefont {N.~I.}\ \bibnamefont
  {Shvetsov-Shilovski}}, \bibinfo {author} {\bibfnamefont {M.}~\bibnamefont
  {Lein}}, \bibinfo {author} {\bibfnamefont {L.~B.}\ \bibnamefont {Madsen}},
  \bibinfo {author} {\bibfnamefont {E.}~\bibnamefont {R\"as\"anen}}, \bibinfo
  {author} {\bibfnamefont {C.}~\bibnamefont {Lemell}}, \bibinfo {author}
  {\bibfnamefont {J.}~\bibnamefont {Burgd\"orfer}}, \bibinfo {author}
  {\bibfnamefont {D.~G.}\ \bibnamefont {Arb\'o}}, \ and\ \bibinfo {author}
  {\bibfnamefont {K.}~\bibnamefont
  {T\ifmmode\mbox{\H{o}}\else\H{o}\fi{}k\'esi}},\ }\href {\doibase
  10.1103/PhysRevA.94.013415} {\bibfield  {journal} {\bibinfo  {journal} {Phys.
  Rev. A}\ }\textbf {\bibinfo {volume} {94}},\ \bibinfo {pages} {013415}
  (\bibinfo {year} {2016})}\BibitemShut {NoStop}%
\bibitem [{\citenamefont {Ni}\ \emph {et~al.}(2016)\citenamefont {Ni},
  \citenamefont {Saalmann},\ and\ \citenamefont {Rost}}]{Ni2016}%
  \BibitemOpen
  \bibfield  {author} {\bibinfo {author} {\bibfnamefont {H.}~\bibnamefont
  {Ni}}, \bibinfo {author} {\bibfnamefont {U.}~\bibnamefont {Saalmann}}, \ and\
  \bibinfo {author} {\bibfnamefont {J.-M.}\ \bibnamefont {Rost}},\ }\href
  {\doibase 10.1103/PhysRevLett.117.023002} {\bibfield  {journal} {\bibinfo
  {journal} {Phys. Rev. Lett.}\ }\textbf {\bibinfo {volume} {117}},\ \bibinfo
  {pages} {023002} (\bibinfo {year} {2016})}\BibitemShut {NoStop}%
\bibitem [{\citenamefont {Ni}\ \emph {et~al.}(2018)\citenamefont {Ni},
  \citenamefont {Saalmann},\ and\ \citenamefont {Rost}}]{Ni2018_theo}%
  \BibitemOpen
  \bibfield  {author} {\bibinfo {author} {\bibfnamefont {H.}~\bibnamefont
  {Ni}}, \bibinfo {author} {\bibfnamefont {U.}~\bibnamefont {Saalmann}}, \ and\
  \bibinfo {author} {\bibfnamefont {J.-M.}\ \bibnamefont {Rost}},\ }\href
  {\doibase 10.1103/PhysRevA.97.013426} {\bibfield  {journal} {\bibinfo
  {journal} {Phys. Rev. A}\ }\textbf {\bibinfo {volume} {97}},\ \bibinfo
  {pages} {013426} (\bibinfo {year} {2018})}\BibitemShut {NoStop}%
\bibitem [{\citenamefont {Shvetsov-Shilovski}\ and\ \citenamefont
  {Lein}(2019)}]{Shilovski2019}%
  \BibitemOpen
  \bibfield  {author} {\bibinfo {author} {\bibfnamefont {N.~I.}\ \bibnamefont
  {Shvetsov-Shilovski}}\ and\ \bibinfo {author} {\bibfnamefont
  {M.}~\bibnamefont {Lein}},\ }\href {\doibase 10.1103/PhysRevA.100.053411}
  {\bibfield  {journal} {\bibinfo  {journal} {Phys. Rev. A}\ }\textbf {\bibinfo
  {volume} {100}},\ \bibinfo {pages} {053411} (\bibinfo {year}
  {2019})}\BibitemShut {NoStop}%
\bibitem [{\citenamefont {Eckle}\ \emph {et~al.}(2008)\citenamefont {Eckle},
  \citenamefont {Pfeiffer}, \citenamefont {Cirelli}, \citenamefont {Staudte},
  \citenamefont {D{\"o}rner}, \citenamefont {Muller}, \citenamefont
  {B{\"u}ttiker},\ and\ \citenamefont {Keller}}]{Eckle2008}%
  \BibitemOpen
  \bibfield  {author} {\bibinfo {author} {\bibfnamefont {P.}~\bibnamefont
  {Eckle}}, \bibinfo {author} {\bibfnamefont {A.~N.}\ \bibnamefont {Pfeiffer}},
  \bibinfo {author} {\bibfnamefont {C.}~\bibnamefont {Cirelli}}, \bibinfo
  {author} {\bibfnamefont {A.}~\bibnamefont {Staudte}}, \bibinfo {author}
  {\bibfnamefont {R.}~\bibnamefont {D{\"o}rner}}, \bibinfo {author}
  {\bibfnamefont {H.~G.}\ \bibnamefont {Muller}}, \bibinfo {author}
  {\bibfnamefont {M.}~\bibnamefont {B{\"u}ttiker}}, \ and\ \bibinfo {author}
  {\bibfnamefont {U.}~\bibnamefont {Keller}},\ }\href {\doibase
  10.1126/science.1163439} {\bibfield  {journal} {\bibinfo  {journal}
  {Science}\ }\textbf {\bibinfo {volume} {322}},\ \bibinfo {pages} {1525}
  (\bibinfo {year} {2008})}\BibitemShut {NoStop}%
\bibitem [{\citenamefont {Arissian}\ \emph {et~al.}(2010)\citenamefont
  {Arissian}, \citenamefont {Smeenk}, \citenamefont {Turner}, \citenamefont
  {Trallero}, \citenamefont {Sokolov}, \citenamefont {Villeneuve},
  \citenamefont {Staudte},\ and\ \citenamefont {Corkum}}]{Arissian2010}%
  \BibitemOpen
  \bibfield  {author} {\bibinfo {author} {\bibfnamefont {L.}~\bibnamefont
  {Arissian}}, \bibinfo {author} {\bibfnamefont {C.}~\bibnamefont {Smeenk}},
  \bibinfo {author} {\bibfnamefont {F.}~\bibnamefont {Turner}}, \bibinfo
  {author} {\bibfnamefont {C.}~\bibnamefont {Trallero}}, \bibinfo {author}
  {\bibfnamefont {A.~V.}\ \bibnamefont {Sokolov}}, \bibinfo {author}
  {\bibfnamefont {D.~M.}\ \bibnamefont {Villeneuve}}, \bibinfo {author}
  {\bibfnamefont {A.}~\bibnamefont {Staudte}}, \ and\ \bibinfo {author}
  {\bibfnamefont {P.~B.}\ \bibnamefont {Corkum}},\ }\href {\doibase
  10.1103/PhysRevLett.105.133002} {\bibfield  {journal} {\bibinfo  {journal}
  {Phys. Rev. Lett.}\ }\textbf {\bibinfo {volume} {105}},\ \bibinfo {pages}
  {133002} (\bibinfo {year} {2010})}\BibitemShut {NoStop}%
\bibitem [{\citenamefont {Barth}\ and\ \citenamefont
  {Smirnova}(2011)}]{Olga2011A}%
  \BibitemOpen
  \bibfield  {author} {\bibinfo {author} {\bibfnamefont {I.}~\bibnamefont
  {Barth}}\ and\ \bibinfo {author} {\bibfnamefont {O.}~\bibnamefont
  {Smirnova}},\ }\href {\doibase 10.1103/PhysRevA.84.063415} {\bibfield
  {journal} {\bibinfo  {journal} {Phys. Rev. A}\ }\textbf {\bibinfo {volume}
  {84}},\ \bibinfo {pages} {063415} (\bibinfo {year} {2011})}\BibitemShut
  {NoStop}%
\bibitem [{\citenamefont {Barth}\ and\ \citenamefont
  {Smirnova}(2013)}]{Olga2011B}%
  \BibitemOpen
  \bibfield  {author} {\bibinfo {author} {\bibfnamefont {I.}~\bibnamefont
  {Barth}}\ and\ \bibinfo {author} {\bibfnamefont {O.}~\bibnamefont
  {Smirnova}},\ }\href {\doibase 10.1103/PhysRevA.87.013433} {\bibfield
  {journal} {\bibinfo  {journal} {Phys. Rev. A}\ }\textbf {\bibinfo {volume}
  {87}},\ \bibinfo {pages} {013433} (\bibinfo {year} {2013})}\BibitemShut
  {NoStop}%
\bibitem [{\citenamefont {Eckart}\ \emph
  {et~al.}(2018{\natexlab{b}})\citenamefont {Eckart}, \citenamefont {Fehre},
  \citenamefont {Eicke}, \citenamefont {Hartung}, \citenamefont {Rist},
  \citenamefont {Trabert}, \citenamefont {Strenger}, \citenamefont {Pier},
  \citenamefont {{{L. Ph. H. Schmidt}}}, \citenamefont {Jahnke}, \citenamefont
  {Sch\"offler}, \citenamefont {Lein}, \citenamefont {Kunitski},\ and\
  \citenamefont {D\"orner}}]{Eckart2018_Offsets}%
  \BibitemOpen
  \bibfield  {author} {\bibinfo {author} {\bibfnamefont {S.}~\bibnamefont
  {Eckart}}, \bibinfo {author} {\bibfnamefont {K.}~\bibnamefont {Fehre}},
  \bibinfo {author} {\bibfnamefont {N.}~\bibnamefont {Eicke}}, \bibinfo
  {author} {\bibfnamefont {A.}~\bibnamefont {Hartung}}, \bibinfo {author}
  {\bibfnamefont {J.}~\bibnamefont {Rist}}, \bibinfo {author} {\bibfnamefont
  {D.}~\bibnamefont {Trabert}}, \bibinfo {author} {\bibfnamefont
  {N.}~\bibnamefont {Strenger}}, \bibinfo {author} {\bibfnamefont
  {A.}~\bibnamefont {Pier}}, \bibinfo {author} {\bibnamefont {{{L. Ph. H.
  Schmidt}}}}, \bibinfo {author} {\bibfnamefont {T.}~\bibnamefont {Jahnke}},
  \bibinfo {author} {\bibfnamefont {M.~S.}\ \bibnamefont {Sch\"offler}},
  \bibinfo {author} {\bibfnamefont {M.}~\bibnamefont {Lein}}, \bibinfo {author}
  {\bibfnamefont {M.}~\bibnamefont {Kunitski}}, \ and\ \bibinfo {author}
  {\bibfnamefont {R.}~\bibnamefont {D\"orner}},\ }\href {\doibase
  10.1103/PhysRevLett.121.163202} {\bibfield  {journal} {\bibinfo  {journal}
  {Phys. Rev. Lett.}\ }\textbf {\bibinfo {volume} {121}},\ \bibinfo {pages}
  {163202} (\bibinfo {year} {2018}{\natexlab{b}})}\BibitemShut {NoStop}%
\bibitem [{\citenamefont {Trabert}\ \emph {et~al.}(2020)\citenamefont
  {Trabert}, \citenamefont {Fehre}, \citenamefont {Anders}, \citenamefont
  {Geyer}, \citenamefont {Grundmann}, \citenamefont {Sch\"offler},
  \citenamefont {{{L. Ph. H. Schmidt}}}, \citenamefont {Jahnke}, \citenamefont
  {D\"orner}, \citenamefont {Kunitski},\ and\ \citenamefont
  {Eckart}}]{DanielArXiv2020}%
  \BibitemOpen
  \bibfield  {author} {\bibinfo {author} {\bibfnamefont {D.}~\bibnamefont
  {Trabert}}, \bibinfo {author} {\bibfnamefont {K.}~\bibnamefont {Fehre}},
  \bibinfo {author} {\bibfnamefont {N.}~\bibnamefont {Anders}}, \bibinfo
  {author} {\bibfnamefont {A.}~\bibnamefont {Geyer}}, \bibinfo {author}
  {\bibfnamefont {S.}~\bibnamefont {Grundmann}}, \bibinfo {author}
  {\bibfnamefont {M.}~\bibnamefont {Sch\"offler}}, \bibinfo {author}
  {\bibnamefont {{{L. Ph. H. Schmidt}}}}, \bibinfo {author} {\bibfnamefont
  {T.}~\bibnamefont {Jahnke}}, \bibinfo {author} {\bibfnamefont
  {R.}~\bibnamefont {D\"orner}}, \bibinfo {author} {\bibfnamefont
  {M.}~\bibnamefont {Kunitski}}, \ and\ \bibinfo {author} {\bibfnamefont
  {S.}~\bibnamefont {Eckart}},\ }\href@noop {} {\bibfield  {journal} {\bibinfo
  {journal} {arXiv preprint, arXiv:2005.09584}\ } (\bibinfo {year}
  {2020})}\BibitemShut {NoStop}%
\bibitem [{\citenamefont {Eckart}\ \emph {et~al.}(2020)\citenamefont {Eckart},
  \citenamefont {Trabert}, \citenamefont {Fehre}, \citenamefont {Geyer},
  \citenamefont {Rist}, \citenamefont {Lin}, \citenamefont {Trinter},
  \citenamefont {{{L. Ph. H. Schmidt}}}, \citenamefont {Sch\"offler},
  \citenamefont {Jahnke}, \citenamefont {Kunitski},\ and\ \citenamefont
  {D\"orner}}]{EckartArXivSideband}%
  \BibitemOpen
  \bibfield  {author} {\bibinfo {author} {\bibfnamefont {S.}~\bibnamefont
  {Eckart}}, \bibinfo {author} {\bibfnamefont {D.}~\bibnamefont {Trabert}},
  \bibinfo {author} {\bibfnamefont {K.}~\bibnamefont {Fehre}}, \bibinfo
  {author} {\bibfnamefont {A.}~\bibnamefont {Geyer}}, \bibinfo {author}
  {\bibfnamefont {J.}~\bibnamefont {Rist}}, \bibinfo {author} {\bibfnamefont
  {K.}~\bibnamefont {Lin}}, \bibinfo {author} {\bibfnamefont {F.}~\bibnamefont
  {Trinter}}, \bibinfo {author} {\bibnamefont {{{L. Ph. H. Schmidt}}}},
  \bibinfo {author} {\bibfnamefont {M.}~\bibnamefont {Sch\"offler}}, \bibinfo
  {author} {\bibfnamefont {T.}~\bibnamefont {Jahnke}}, \bibinfo {author}
  {\bibfnamefont {M.}~\bibnamefont {Kunitski}}, \ and\ \bibinfo {author}
  {\bibfnamefont {R.}~\bibnamefont {D\"orner}},\ }\href@noop {} {\bibfield
  {journal} {\bibinfo  {journal} {arXiv preprint, arXiv:2005.04148}\ }
  (\bibinfo {year} {2020})}\BibitemShut {NoStop}%
\bibitem [{\citenamefont {Yudin}\ and\ \citenamefont
  {Ivanov}(2001)}]{Yudin2001A}%
  \BibitemOpen
  \bibfield  {author} {\bibinfo {author} {\bibfnamefont {G.~L.}\ \bibnamefont
  {Yudin}}\ and\ \bibinfo {author} {\bibfnamefont {M.~Y.}~\bibnamefont {Ivanov}},\
  }\href {\doibase 10.1103/PhysRevA.64.013409} {\bibfield  {journal} {\bibinfo
  {journal} {Phys. Rev. A}\ }\textbf {\bibinfo {volume} {64}},\ \bibinfo
  {pages} {013409} (\bibinfo {year} {2001})}\BibitemShut {NoStop}%
\bibitem [{\citenamefont {Torlina}\ \emph {et~al.}(2015)\citenamefont
  {Torlina}, \citenamefont {Morales}, \citenamefont {Kaushal}, \citenamefont
  {Ivanov}, \citenamefont {Kheifets}, \citenamefont {Zielinski}, \citenamefont
  {Scrinzi}, \citenamefont {Muller}, \citenamefont {Sukiasyan}, \citenamefont
  {Ivanov},\ and\ \citenamefont {Olga}}]{torlina2015interpreting}%
  \BibitemOpen
  \bibfield  {author} {\bibinfo {author} {\bibfnamefont {L.}~\bibnamefont
  {Torlina}}, \bibinfo {author} {\bibfnamefont {F.}~\bibnamefont {Morales}},
  \bibinfo {author} {\bibfnamefont {J.}~\bibnamefont {Kaushal}}, \bibinfo
  {author} {\bibfnamefont {I.}~\bibnamefont {Ivanov}}, \bibinfo {author}
  {\bibfnamefont {A.}~\bibnamefont {Kheifets}}, \bibinfo {author}
  {\bibfnamefont {A.}~\bibnamefont {Zielinski}}, \bibinfo {author}
  {\bibfnamefont {A.}~\bibnamefont {Scrinzi}}, \bibinfo {author} {\bibfnamefont
  {H.~G.}\ \bibnamefont {Muller}}, \bibinfo {author} {\bibfnamefont
  {S.}~\bibnamefont {Sukiasyan}}, \bibinfo {author} {\bibfnamefont
  {M.}~\bibnamefont {Ivanov}}, \ and\ \bibinfo {author} {\bibfnamefont
  {S.}~\bibnamefont {Olga}},\ }\href
  {https://www.nature.com/articles/nphys3340} {\bibfield  {journal} {\bibinfo
  {journal} {Nat. Phys.}\ }\textbf {\bibinfo {volume} {11}},\ \bibinfo {pages}
  {503} (\bibinfo {year} {2015})}\BibitemShut {NoStop}%
\bibitem [{\citenamefont {Bray}\ \emph {et~al.}(2018)\citenamefont {Bray},
  \citenamefont {Eckart},\ and\ \citenamefont {Kheifets}}]{Bray2018}%
  \BibitemOpen
  \bibfield  {author} {\bibinfo {author} {\bibfnamefont {A.~W.}\ \bibnamefont
  {Bray}}, \bibinfo {author} {\bibfnamefont {S.}~\bibnamefont {Eckart}}, \ and\
  \bibinfo {author} {\bibfnamefont {A.~S.}\ \bibnamefont {Kheifets}},\ }\href
  {\doibase 10.1103/PhysRevLett.121.123201} {\bibfield  {journal} {\bibinfo
  {journal} {Phys. Rev. Lett.}\ }\textbf {\bibinfo {volume} {121}},\ \bibinfo
  {pages} {123201} (\bibinfo {year} {2018})}\BibitemShut {NoStop}%
\bibitem [{\citenamefont {Kunitski}\ \emph {et~al.}(2019)\citenamefont
  {Kunitski}, \citenamefont {Eicke}, \citenamefont {Huber}, \citenamefont
  {K{\"o}hler}, \citenamefont {Zeller}, \citenamefont {Voigtsberger},
  \citenamefont {Schlott}, \citenamefont {Henrichs}, \citenamefont {Sann},
  \citenamefont {Trinter}, \citenamefont {{{L. Ph. H. Schmidt}}}, \citenamefont
  {Kalinin}, \citenamefont {Sch\"offler}, \citenamefont {Jahnke}, \citenamefont
  {Lein},\ and\ \citenamefont {D\"orner}}]{MaksimNatureCom}%
  \BibitemOpen
  \bibfield  {author} {\bibinfo {author} {\bibfnamefont {M.}~\bibnamefont
  {Kunitski}}, \bibinfo {author} {\bibfnamefont {N.}~\bibnamefont {Eicke}},
  \bibinfo {author} {\bibfnamefont {P.}~\bibnamefont {Huber}}, \bibinfo
  {author} {\bibfnamefont {J.}~\bibnamefont {K{\"o}hler}}, \bibinfo {author}
  {\bibfnamefont {S.}~\bibnamefont {Zeller}}, \bibinfo {author} {\bibfnamefont
  {J.}~\bibnamefont {Voigtsberger}}, \bibinfo {author} {\bibfnamefont
  {N.}~\bibnamefont {Schlott}}, \bibinfo {author} {\bibfnamefont
  {K.}~\bibnamefont {Henrichs}}, \bibinfo {author} {\bibfnamefont
  {H.}~\bibnamefont {Sann}}, \bibinfo {author} {\bibfnamefont {F.}~\bibnamefont
  {Trinter}}, \bibinfo {author} {\bibnamefont {{{L. Ph. H. Schmidt}}}},
  \bibinfo {author} {\bibfnamefont {A.}~\bibnamefont {Kalinin}}, \bibinfo
  {author} {\bibfnamefont {M.~S.}\ \bibnamefont {Sch\"offler}}, \bibinfo
  {author} {\bibfnamefont {T.}~\bibnamefont {Jahnke}}, \bibinfo {author}
  {\bibfnamefont {M.}~\bibnamefont {Lein}}, \ and\ \bibinfo {author}
  {\bibfnamefont {R.}~\bibnamefont {D\"orner}},\ }\href@noop {} {\bibfield
  {journal} {\bibinfo  {journal} {Nat. Commun}\ }\textbf {\bibinfo {volume}
  {10}},\ \bibinfo {pages} {1} (\bibinfo {year} {2019})}\BibitemShut {NoStop}%
\bibitem [{\citenamefont {Fechner}\ \emph {et~al.}(2014)\citenamefont
  {Fechner}, \citenamefont {Camus}, \citenamefont {Ullrich}, \citenamefont
  {Pfeifer},\ and\ \citenamefont {Moshammer}}]{Fechner2014}%
  \BibitemOpen
  \bibfield  {author} {\bibinfo {author} {\bibfnamefont {L.}~\bibnamefont
  {Fechner}}, \bibinfo {author} {\bibfnamefont {N.}~\bibnamefont {Camus}},
  \bibinfo {author} {\bibfnamefont {J.}~\bibnamefont {Ullrich}}, \bibinfo
  {author} {\bibfnamefont {T.}~\bibnamefont {Pfeifer}}, \ and\ \bibinfo
  {author} {\bibfnamefont {R.}~\bibnamefont {Moshammer}},\ }\href {\doibase
  10.1103/PhysRevLett.112.213001} {\bibfield  {journal} {\bibinfo  {journal}
  {Phys. Rev. Lett.}\ }\textbf {\bibinfo {volume} {112}},\ \bibinfo {pages}
  {213001} (\bibinfo {year} {2014})}\BibitemShut {NoStop}%
\bibitem [{\citenamefont {Meckel}\ \emph {et~al.}(2008)\citenamefont {Meckel},
  \citenamefont {Comtois}, \citenamefont {Zeidler}, \citenamefont {Staudte},
  \citenamefont {Pavi{\v c}i{\'c}}, \citenamefont {Bandulet}, \citenamefont
  {P{\'e}pin}, \citenamefont {Kieffer}, \citenamefont {D{\"o}rner},
  \citenamefont {Villeneuve},\ and\ \citenamefont {Corkum}}]{Meckel2008}%
  \BibitemOpen
  \bibfield  {author} {\bibinfo {author} {\bibfnamefont {M.}~\bibnamefont
  {Meckel}}, \bibinfo {author} {\bibfnamefont {D.}~\bibnamefont {Comtois}},
  \bibinfo {author} {\bibfnamefont {D.}~\bibnamefont {Zeidler}}, \bibinfo
  {author} {\bibfnamefont {A.}~\bibnamefont {Staudte}}, \bibinfo {author}
  {\bibfnamefont {D.}~\bibnamefont {Pavi{\v c}i{\'c}}}, \bibinfo {author}
  {\bibfnamefont {H.~C.}\ \bibnamefont {Bandulet}}, \bibinfo {author}
  {\bibfnamefont {H.}~\bibnamefont {P{\'e}pin}}, \bibinfo {author}
  {\bibfnamefont {J.~C.}\ \bibnamefont {Kieffer}}, \bibinfo {author}
  {\bibfnamefont {R.}~\bibnamefont {D{\"o}rner}}, \bibinfo {author}
  {\bibfnamefont {D.~M.}\ \bibnamefont {Villeneuve}}, \ and\ \bibinfo {author}
  {\bibfnamefont {P.~B.}\ \bibnamefont {Corkum}},\ }\href {\doibase
  10.1126/science.1157980} {\bibfield  {journal} {\bibinfo  {journal}
  {Science}\ }\textbf {\bibinfo {volume} {320}},\ \bibinfo {pages} {1478}
  (\bibinfo {year} {2008})}\BibitemShut {NoStop}%
\bibitem [{\citenamefont {Eckart}\ \emph
  {et~al.}(2018{\natexlab{c}})\citenamefont {Eckart}, \citenamefont {Kunitski},
  \citenamefont {Richter}, \citenamefont {Hartung}, \citenamefont {Rist},
  \citenamefont {Trinter}, \citenamefont {Fehre}, \citenamefont {Schlott},
  \citenamefont {Henrichs}, \citenamefont {{{L. Ph. H. Schmidt}}},
  \citenamefont {Jahnke}, \citenamefont {Sch{\"{o}}ffler}, \citenamefont {Liu},
  \citenamefont {Barth}, \citenamefont {Kaushal}, \citenamefont {Morales},
  \citenamefont {Ivanov}, \citenamefont {Smirnova},\ and\ \citenamefont
  {D{\"{o}}rner}}]{EckartNatPhys2018}%
  \BibitemOpen
  \bibfield  {author} {\bibinfo {author} {\bibfnamefont {S.}~\bibnamefont
  {Eckart}}, \bibinfo {author} {\bibfnamefont {M.}~\bibnamefont {Kunitski}},
  \bibinfo {author} {\bibfnamefont {M.}~\bibnamefont {Richter}}, \bibinfo
  {author} {\bibfnamefont {A.}~\bibnamefont {Hartung}}, \bibinfo {author}
  {\bibfnamefont {J.}~\bibnamefont {Rist}}, \bibinfo {author} {\bibfnamefont
  {F.}~\bibnamefont {Trinter}}, \bibinfo {author} {\bibfnamefont
  {K.}~\bibnamefont {Fehre}}, \bibinfo {author} {\bibfnamefont
  {N.}~\bibnamefont {Schlott}}, \bibinfo {author} {\bibfnamefont
  {K.}~\bibnamefont {Henrichs}}, \bibinfo {author} {\bibnamefont {{{L. Ph. H.
  Schmidt}}}}, \bibinfo {author} {\bibfnamefont {T.}~\bibnamefont {Jahnke}},
  \bibinfo {author} {\bibfnamefont {M.}~\bibnamefont {Sch{\"{o}}ffler}},
  \bibinfo {author} {\bibfnamefont {K.}~\bibnamefont {Liu}}, \bibinfo {author}
  {\bibfnamefont {I.}~\bibnamefont {Barth}}, \bibinfo {author} {\bibfnamefont
  {J.}~\bibnamefont {Kaushal}}, \bibinfo {author} {\bibfnamefont
  {F.}~\bibnamefont {Morales}}, \bibinfo {author} {\bibfnamefont
  {M.}~\bibnamefont {Ivanov}}, \bibinfo {author} {\bibfnamefont
  {O.}~\bibnamefont {Smirnova}}, \ and\ \bibinfo {author} {\bibfnamefont
  {R.}~\bibnamefont {D{\"{o}}rner}},\ }\href {\doibase
  10.1038/s41567-018-0080-5} {\bibfield  {journal} {\bibinfo  {journal} {Nat.
  Phys.}\ }\textbf {\bibinfo {volume} {14}},\ \bibinfo {pages} {701} (\bibinfo
  {year} {2018}{\natexlab{c}})}\BibitemShut {NoStop}%
\bibitem [{\citenamefont {Vos}\ \emph {et~al.}(2018)\citenamefont {Vos},
  \citenamefont {Cattaneo}, \citenamefont {Patchkovskii}, \citenamefont
  {Zimmermann}, \citenamefont {Cirelli}, \citenamefont {Lucchini},
  \citenamefont {Kheifets}, \citenamefont {Landsman},\ and\ \citenamefont
  {Keller}}]{vos2018orientation}%
  \BibitemOpen
  \bibfield  {author} {\bibinfo {author} {\bibfnamefont {J.}~\bibnamefont
  {Vos}}, \bibinfo {author} {\bibfnamefont {L.}~\bibnamefont {Cattaneo}},
  \bibinfo {author} {\bibfnamefont {S.}~\bibnamefont {Patchkovskii}}, \bibinfo
  {author} {\bibfnamefont {T.}~\bibnamefont {Zimmermann}}, \bibinfo {author}
  {\bibfnamefont {C.}~\bibnamefont {Cirelli}}, \bibinfo {author} {\bibfnamefont
  {M.}~\bibnamefont {Lucchini}}, \bibinfo {author} {\bibfnamefont
  {A.}~\bibnamefont {Kheifets}}, \bibinfo {author} {\bibfnamefont {A.~S.}\
  \bibnamefont {Landsman}}, \ and\ \bibinfo {author} {\bibfnamefont
  {U.}~\bibnamefont {Keller}},\ }\href@noop {} {\bibfield  {journal} {\bibinfo
  {journal} {Science}\ }\textbf {\bibinfo {volume} {360}},\ \bibinfo {pages}
  {1326} (\bibinfo {year} {2018})}\BibitemShut {NoStop}%
\bibitem [{\citenamefont {Paul}\ \emph {et~al.}(2001)\citenamefont {Paul},
  \citenamefont {Toma}, \citenamefont {Breger}, \citenamefont {Mullot},
  \citenamefont {Aug{\'e}}, \citenamefont {Balcou}, \citenamefont {Muller},\
  and\ \citenamefont {Agostini}}]{Paul1689}%
  \BibitemOpen
  \bibfield  {author} {\bibinfo {author} {\bibfnamefont {P.~M.}\ \bibnamefont
  {Paul}}, \bibinfo {author} {\bibfnamefont {E.~S.}\ \bibnamefont {Toma}},
  \bibinfo {author} {\bibfnamefont {P.}~\bibnamefont {Breger}}, \bibinfo
  {author} {\bibfnamefont {G.}~\bibnamefont {Mullot}}, \bibinfo {author}
  {\bibfnamefont {F.}~\bibnamefont {Aug{\'e}}}, \bibinfo {author}
  {\bibfnamefont {P.}~\bibnamefont {Balcou}}, \bibinfo {author} {\bibfnamefont
  {H.~G.}\ \bibnamefont {Muller}}, \ and\ \bibinfo {author} {\bibfnamefont
  {P.}~\bibnamefont {Agostini}},\ }\href {\doibase 10.1126/science.1059413}
  {\bibfield  {journal} {\bibinfo  {journal} {Science}\ }\textbf {\bibinfo
  {volume} {292}},\ \bibinfo {pages} {1689} (\bibinfo {year}
  {2001})}\BibitemShut {NoStop}%
\bibitem [{\citenamefont {Muller}(2002)}]{Muller2002}%
  \BibitemOpen
  \bibfield  {author} {\bibinfo {author} {\bibfnamefont {H.}~\bibnamefont
  {Muller}},\ }\href {\doibase 10.1007/s00340-002-0894-8} {\bibfield  {journal}
  {\bibinfo  {journal} {Applied Physics B}\ }\textbf {\bibinfo {volume} {74}},\
  \bibinfo {pages} {s17} (\bibinfo {year} {2002})}\BibitemShut {NoStop}%
\bibitem [{\citenamefont {de~Carvalho}\ and\ \citenamefont
  {Nussenzveig}(2002)}]{Carvalho2002}%
  \BibitemOpen
  \bibfield  {author} {\bibinfo {author} {\bibfnamefont {C.~A.~A.}\
  \bibnamefont {de~Carvalho}}\ and\ \bibinfo {author} {\bibfnamefont {H.~M.}\
  \bibnamefont {Nussenzveig}},\ }\href@noop {} {\bibfield  {journal} {\bibinfo
  {journal} {Physics Reports}\ }\textbf {\bibinfo {volume} {364}},\ \bibinfo
  {pages} {83} (\bibinfo {year} {2002})}\BibitemShut {NoStop}%
\bibitem [{\citenamefont {Ivanov}\ and\ \citenamefont
  {Kheifets}(2013)}]{Ivanov2013}%
  \BibitemOpen
  \bibfield  {author} {\bibinfo {author} {\bibfnamefont {I.~A.}\ \bibnamefont
  {Ivanov}}\ and\ \bibinfo {author} {\bibfnamefont {A.~S.}\ \bibnamefont
  {Kheifets}},\ }\href {\doibase 10.1103/PhysRevA.87.033407} {\bibfield
  {journal} {\bibinfo  {journal} {Phys. Rev. A}\ }\textbf {\bibinfo {volume}
  {87}},\ \bibinfo {pages} {033407} (\bibinfo {year} {2013})}\BibitemShut
  {NoStop}%
\bibitem [{\citenamefont {Cirelli}\ \emph {et~al.}(2015)\citenamefont
  {Cirelli}, \citenamefont {Heuser}, \citenamefont {Boge}, \citenamefont
  {Lucchini}, \citenamefont {Gallmann},\ and\ \citenamefont
  {Keller}}]{cirelli2015energy}%
  \BibitemOpen
  \bibfield  {author} {\bibinfo {author} {\bibfnamefont {M.}~\bibnamefont
  {Cirelli}, \bibfnamefont {C.and~Sabbar}}, \bibinfo {author} {\bibfnamefont
  {S.}~\bibnamefont {Heuser}}, \bibinfo {author} {\bibfnamefont
  {R.}~\bibnamefont {Boge}}, \bibinfo {author} {\bibfnamefont {M.}~\bibnamefont
  {Lucchini}}, \bibinfo {author} {\bibfnamefont {L.}~\bibnamefont {Gallmann}},
  \ and\ \bibinfo {author} {\bibfnamefont {U.}~\bibnamefont {Keller}},\
  }\href@noop {} {\bibfield  {journal} {\bibinfo  {journal} {IEEE Journal of
  Selected Topics in Quantum Electronics}\ }\textbf {\bibinfo {volume} {21}},\
  \bibinfo {pages} {1} (\bibinfo {year} {2015})}\BibitemShut {NoStop}%
\bibitem [{Note1()}]{Note1}%
  \BibitemOpen
  \bibinfo {note} {The tunneling probability for the SCTS model is assumed to
  be $R(\protect \mathaccentV {vec}17E{p}_{\protect \mathrm {i\protect
  \_3D}})=\protect \qopname \relax o{exp}\left (-\protect \frac {\left
  |\protect \mathaccentV {vec}17E{p}_{\protect \mathrm {i\protect
  \_3D}}-\protect \mathaccentV {vec}17E{p}_{\protect \mathrm {0\protect
  \_3D}}\right |^2}{2\sigma ^2}\right )$, which is in full analogy to Fig. \ref
  {fig_linearphasemomdistribution}(b). The initial momentum at the tunnel exit
  is represented by $\protect \mathaccentV {vec}17E{p}_{\protect \mathrm
  {i\protect \_3D}}=\begin {pmatrix}p_{\protect \mathrm {0x}}\\p_{\protect
  \mathrm {0\perp }}\\p_{\protect \mathrm {0\parallel }}\end {pmatrix}$. Here,
  $\protect \mathaccentV {vec}17E{p}_{\protect \mathrm {i\protect \_3D}}$ is
  defined using a reference frame that is aligned along the direction of the
  electric field, $\protect \mathaccentV {vec}17E{E}(t_0)$, at the time the
  electron is released, $t_0$. Accordingly, $p_{\protect \mathrm {0x}}$ points
  along the light propagation direction and $p_{\protect \mathrm {0\parallel
  }}$ [$p_{\protect \mathrm {0\perp }}$] points along the direction that is
  parallel (perpendicular) to $\protect \mathaccentV {vec}17E{E}(t_0)$. The
  alignment of the direction that belongs to $p_{\protect \mathrm {0\perp }}$
  is chosen such that an increase in the value of $p_{\protect \mathrm {0\perp
  }}$ leads to an increased absolute value of the final electron momentum (as
  for the HASE model, see discussion of Eq. \ref {lab_initial}). Here, we
  choose $\sigma =0.2$\protect \tmspace +\thinmuskip {.1667em}a.u.,
  $p_{\protect \mathrm {0\parallel }}=0$\protect \tmspace +\thinmuskip
  {.1667em}a.u. and $\protect \mathaccentV {vec}17E{p}_{\protect \mathrm
  {0\protect \_3D}}=\begin {pmatrix}0\\0.2\\0\end {pmatrix}$ in full analogy to
  the HASE model. It should be noted, that due to the choice of $R(\protect
  \mathaccentV {vec}17E{p}_{\protect \mathrm {i\protect \_3D}})$, the tunneling
  probability, does not depend on the absolute value of the electric field,
  which is a difference compared to the original SCTS model (see Eq. 9 from
  Ref. \cite {Shilovski2016}).}\BibitemShut {Stop}%
\bibitem [{\citenamefont {Eckart}(2019)}]{EckartDiss}%
  \BibitemOpen
  \bibfield  {author} {\bibinfo {author} {\bibfnamefont {S.~G.}\ \bibnamefont
  {Eckart}},\ }\href@noop {} {\bibfield  {journal} {\bibinfo  {journal}
  {Ph.D. thesis, Goethe University Frankfurt, Frankfurt am Main (Germany)}\ }
  (\bibinfo {year} {2019})}\BibitemShut {NoStop}%
\end{thebibliography}

\end{document}